\documentclass[preprint,showpacs,showkeys,preprintnumbers,amsmath,amssymb]{revtex4}

\usepackage{graphicx}
\usepackage{dcolumn}
\usepackage{bm}

\begin{document}

\preprint{}

\title{Hilbert-Schmidt Geometry of 
$n$-Level Jak\'obczyk-Siennicki Two-Dimensional Quantum Systems}

\author{Paul B. Slater}%
\email{slater@kitp.ucsb.edu}
\affiliation{%
ISBER, University of California, Santa Barbara, CA 93106\\
}%
\date{\today}

\begin{abstract}
Jak\'obczyk and Siennicki studied {\it two}-dimensional sections of a set of
(generalized) 
Bloch vectors corresponding to $n \times n$  density matrices of two-qubit systems (that is, the case $n=4$).
They found essentially five different types of (nontrivial) separability regimes.
We compute 
the Euclidean/Hilbert-Schmidt 
(HS) separability {\it probabilities} assigned to these
regimes, and conduct parallel {\it two}-dimensional sectional
analyses for the 
{\it higher}-level cases $n=6,8,9$ and 10. Making use of the newly-introduced
capability for  integration over {\it implicitly} defined regions of version 5.1 of 
Mathematica --- as we have also fruitfully done in 
the $n=4$ {\it three}-parameter 
entropy-maximization-based study quant-ph/0507203 --- we 
obtain a wide-ranging variety of 
{\it exact} 
HS-probabilities.
For $n>6$, the probabilities are those of having a {\it partial positive
transpose} (PPT). For the $n=6$ case, we also obtain {\it biseparability} 
probabilities; in the $n=8,9 $ instances, 
bi-PPT  probabilities; and for $n=8$, {\it tri}-PPT probabilities.
By far, the most frequently recorded
probability for $n>4$ is  $\frac{\pi}{4}
\approx 0.785398$. We also conduct a number of related analyses, pertaining 
to the (one-dimensional) {\it boundaries} (both 
exterior and interior) of the separability and PPT domains, and attempt 
(with quite limited success) some
 {\it exact} 
calculations pertaining to the {\it 9}-dimensional (real) and {\it 15}-dimensional
(complex) convex sets of two-qubit density matrices --- for which exact
HS-separability probabilities have been conjectured, but not yet verified.
\end{abstract}

\pacs{Valid PACS 02.40.Dr, 02.40.Ft, 03.67.-a}
\keywords{separability probabilities,  Hilbert-Schmidt metric,  density matrices, Bloch vectors, positive partial transpose,  two-dimensional sections}

\maketitle
\section{Introduction}
There has been considerable recent interest \cite{kk,kimura,byrd} 
in understanding how one can, from the spherical coordinate point-of view, 
 generalize to $n$-level
quantum systems ($n \ge 2$) the familiar Bloch ball representation of the
two-level quantum systems ($n=2$) --- in which the pure states form the
bounding spherical surface (``Bloch sphere'') of the unit ball in
three-dimensional Euclidean space.
Kimura and Kossakowski have expressed the generalized Bloch representation of
an $n \times n$ density matrix in the form \cite[eq. (3)]{kk}
\begin{equation} \label{expansion}
\rho = \frac{\mbox{tr} \rho}{n} I_{n} +\frac{1}{2} \Sigma_{i=1}^{n^2-1} (\mbox{tr} \rho \lambda_{i}) \lambda_{i},
\end{equation}
where $I_{n}$ is the identity operator, and the $\lambda_{i}$'s are the
$(n^2-1)$ 
orthogonal generators of $SU(n)$, forming a basis of the set of all the 
linear operators with respect to the Hilbert-Schmidt inner product.

An interesting application of these concepts was made by 
Jak\'obczyk and Siennicki (JS) \cite{jak}. They  examined all those 
two-qubit ($n=4$) 
systems describable as {\it two}-dimensional sections of sets of 
(generalized) 
Bloch (coherence \cite{byrd}) 
vectors. (The {\it totality} of $4 \times 4$ density matrices, on the other hand, 
 comprises a {\it fifteen}-dimensional convex set --- 
the $n \times n$ density matrices being $(n^2-1)$-dimensional 
in nature --- so thirteen of the
fifteen $SU(4)$ orthogonal generators [Gell-mann matrices] are assigned null 
weight in the JS $n = 4$ analyses. That is, thirteen of the fifteen 
coefficients, $(\mbox{tr} \lambda_{i})$ in the expansion (\ref{expansion}) are
zero.) 

Since there were only two parameters involved in each of their scenarios, 
JS were able to present {\it planar} 
diagrams depicting the feasible regions, as well as those subsets of these regions composed of separable states. 
In their Fig. 1, JS exhibited  thirteen possible types of parameter domains.
Further, in their Fig. 2, they showed  six different  (nontrivial) separability
scenarios (two of which --- labelled ``EF)'' and ``FE)'' by JS -- are simply 
geometric 
reflections of one another). 

We will, firstly (sec.~\ref{secN=4}), in this study, evaluate the sizes 
(areas) 
of these six  (two-dimensional) domains and nontrivial subdomains, 
in terms of the {\it Hilbert-Schmidt} (HS)  metric --- a task JS did 
not explicitly address. 
(The HS-distance between two density operators $\rho_{1},\rho_{2}$ is
defined as $\sqrt{\mbox{Tr}(\rho_{1}-\rho_{2})^2}$ \cite[eq. (2.3)]{hilb2}.)
Then, we extend the JS analyses to the cases $n=6$ (sec.~\ref{secN=6}), 
8 (sec.~\ref{secN=8}), 9 (sec.~\ref{secN=9}) and 10 (sec.~\ref{secN=10}), in 
which various {\it multi}partite --- as opposed to simply {\it bi}partite 
scenarios can arise.
For all these instances, 
except $n=10$, we 
additionally obtain the HS-lengths of the (one-dimensional) boundary
states (that is, those with {\it degenerate} spectra) and the HS-probabilities
that states lying on this boundary are separable.
(Motivated by our extensive numerical 
results given in \cite{slaterPRA} and \cite{ten}, Szarek, Bengtsson 
and \.Zyczkowski have recently 
 proved ``that the probability to find a random state to be separable equals 2 times the probability to find a random boundary state to be separable, provided the random states are generated uniformly with respect to the Hilbert-Schmidt (Euclidean) distance'' \cite{sbz}).
Also, we compute in certain cases, the HS-lengths of the (interior)
boundaries dividing one domain of interest from another. (The interior states 
generically 
have {\it nondegenerate} spectra.) Then (sec.~\ref{finalSEC}), we undertake some analyses
involving {\it three} (rather than two) parameters. These prove to be much more
{\it problematical} in nature (cf. \cite{slaterjpanew}). 

We also report, at the end (sec.~\ref{final}), 
some initial steps in an attempt to determine
 exact {\it upper} bounds for the
HS-volumes of the {\it separable} 
{\it 9}-dimensional real and {\it 15}-dimensional complex
$4 \times 4$ density matrices. 
(Only in these 
computations --- in order to compare our formulas
with known HS-volumes of separable and nonseparable steps \cite{hilb2} --- do
 we not take the HS-volume element to be {\it unity}.)

Our computations in this paper were {\it greatly} 
facilitated by a new feature of the programming 
language Mathematica (version 
5.1) --- the capacity 
to integrate over {\it implicitly} defined regions. 
(This feature was also employed by us in \cite{slaterjpanew}, in a 
somewhat related two-qubit context, in which the Jaynes maximum-entropy 
principle was employed.) We, first, 
found explicit forms for the 
$n$ eigenvalues of the various $n \times n$ 
matrices {\it and} for their partial transposes. 
Then we required, in the several integrations, 
using the new feature  (thus, 
saving us from the laborious task of having to specify
large numbers of particular 
integration limits and do 
corresponding detailed bookkeeping),  simply that these eigenvalues be {\it 
nonnegative}. This ensured that we either had, in fact, 
the requisite {\it density} matrices and/or
{\it positive partial transposes}  (PPT) of density matrices.

\section{The qubit-qubit case $n=4$ of Jak{\o}bczyk and Siennicki} 
\label{secN=4}
 For the  (geometrically-reflected) scenarios that JS labeled 
 ``EF)'' and ``FE)'', we have found 
(Table~\ref{tab:tabJS1})
that the Hilbert-Schmidt 
volume (cf. \cite{hilb2}) of separable {\it and} nonseparable 
states is $\frac{2 \sqrt{2}}{3}$ and
of the separable states {\it alone} is $\frac{2}{3}$. So the 
corresponding separability 
{\it probability} (taking ratios) is --- elegantly --- $\frac{1}{\sqrt{2}} 
\approx 0.707107$.

For the 
scenario ``CK)'', possessing a triangular separability domain, 
the total volume is $\frac{4 \sqrt{\frac{2}{3}}}{3}$
and the separability probability is $\frac{1}{24} (9 +2 \sqrt{3} \pi) 
\approx 0.828450$. For ``GH')'', the total volume is 
   $\frac{9}{32} \sqrt{\frac{3}{2}} \pi$ and the separability probability,
$\frac{26 \sqrt{2}+27 \tan ^{-1}\left(2 \sqrt{2}\right)}{27
   \pi }
\approx 0.825312$. For ``KC)'', the total volume is $\frac{\sqrt{2} \pi }{3}$
and the separability probability is (the smallest)  
$\frac{1}{3} +\frac{\sqrt{3}}{2 \pi}
\approx 0.608998$. For ``HG')'', the total HS-volume is 
$\frac{3}{2 \sqrt{2}}$ and the HS-separability probability is 
(the largest of the five) 
$\frac{52+27 \sqrt{2} \sec ^{-1}(3)}{48 \sqrt{6}} \approx 0.842035$.

\begin{table}[ht]
\caption{\label{tab:tabJS1}}
\begin{ruledtabular}
\begin{tabular}{rrrrr}
\hline
JS scenario  & HS total vol.  &  HS separable  vol.  & HS sep. prob.  & num. approx. \\
\hline
EF) and FE) & $ \frac{2 \sqrt{2}}{3}$  & $ \frac{2}{3}$  & $\frac{1}{{\sqrt{2}}}$ &  0.707107 \\
CK) & $\frac{4 \sqrt{\frac{2}{3}}}{3} $ &
$\frac{1}{18} \left(3 \sqrt{6}+2 \sqrt{2} \pi \right) $ &
$\frac{9 + 2\, {\sqrt{3}}\, \pi}{24} $ &  0.828450 \\
GH') &  $\frac{9}{32} \sqrt{\frac{3}{2}} \pi$ & $ \frac{1}{192} \left(52 \sqrt{3}+27 \sqrt{6} \sin
   ^{-1}\left(\frac{2 \sqrt{2}}{3}\right)\right)$ & $ \frac{26 \sqrt{2}+27 \tan ^{-1}\left(2 \sqrt{2}\right)}{27
   \pi }$ & 
  0.825312 \\
HG') & $\frac{3}{2 \sqrt{2}}$ & 
 $ \frac{1}{192} \left(52 \sqrt{3}+27 \sqrt{6} \sec
   ^{-1}(3)\right)$ & 
 $ \frac{52+27 \sqrt{2} \sec ^{-1}(3)}{48 \sqrt{6}}$ & 0.842035 \\
KC) & $\frac{\sqrt{2} \pi }{3}$ & $ \frac{1}{18} \left(3 \sqrt{6}+2 \sqrt{2} \pi \right)$ & $ \frac{1}{3}+\frac{\sqrt{3}}{2 \pi }$ & 0.608998 \\
\end{tabular}
\end{ruledtabular}
\end{table}

Now, let us present again most of these results (Table~\ref{tab:tabJS1})
in the form of the array (\ref{n=4case}). 
We do so because we will also present all the results of our subsequent 
analyses below (for $n>4$) in this manner (which we have found to be 
the most 
convenient for directly incorporating
 our large-scale Mathematica computer-generated analyses into this report).

In the first column of (\ref{n=4case}) are given  the 
identifying 
numbers of a {\it pair} of Gell-Mann matrices (generators of $SU(4)$) --- which, in fact,  can be seen to fully
agree  with the numbering (and associated scenario-labelling) 
of JS \cite[p. 389]{jak}. 
(Here and further, we will {\it always} adhere to the conventional/standard numbering 
\cite[sec. III]{todd} of the Lie generators
of $SU(n)$, so that our results should be reproducible/verifiable 
to others. 
We list the pairs in lexographic order, using the first pair as
the representative for its equivalence class.)
In the second column of (\ref{n=4case}) 
are shown the {\it number} of distinct unordered pairs
of $SU(4)$ generators which share the same total (separable and nonseparable)
HS volume, as well as the same separable HS volume, and consequently,
identical HS separability probabilities. The third column gives us these
HS total volumes, the fourth column, the HS separability probabilities and 
the last (fifth) column, numerical approximations to the exact probabilities 
(which, of course, we see --- being probabilities --- do not exceed the  value 1).
 (Due to space/page width constraints, we were unable to generally present
in these data arrays 
the HS separable volumes too, 
though they can, of course, be deduced from the total volume and the separability probability.)
\begin{equation} \label{n=4case}
\left(
\begin{array}{lllll}
 \{3,6\} & 4 & \frac{2 \sqrt{2}}{3} & \frac{1}{\sqrt{2}} &
   0.707107 \\
 \{6,8\} & 2 & \frac{9}{32} \sqrt{\frac{3}{2}} \pi  &
   \frac{26 \sqrt{2}+27 \tan ^{-1}\left(2
   \sqrt{2}\right)}{27 \pi } & 0.825312 \\
 \{6,15\} & 2 & \frac{4 \sqrt{\frac{2}{3}}}{3} &
   \frac{1}{24} \left(9+2 \sqrt{3} \pi \right) & 0.828450
   \\
 \{8,9\} & 2 & \frac{3}{2 \sqrt{2}} & \frac{52+27 \sqrt{2}
   \sec ^{-1}(3)}{48 \sqrt{6}} & 0.842035 \\
 \{9,15\} & 2 & \frac{\sqrt{2} \pi }{3} &
   \frac{1}{3}+\frac{\sqrt{3}}{2 \pi } & 0.608998
\end{array}
\right).
\end{equation}
Thus, twelve of the $210 =15 \cdot 14$ possible unordered pairs of
Gell-Mann matrices are associated with nontrivial ($ <  1$) separability
probabilities \cite[p. 389]{jak}.

\subsection{Boundary states}
For the scenario associated with the pair of Gell-Mann matrices
$\{3,6 \}$, the HS-length of
the (one-dimensional) {\it boundary} states  (that is, those with {\it degenerate} spectra) 
is $\frac{3}{2}$, and of the bounding states which are separable, 
$\frac{1}{2}$. 
For the pair $\{6,8 \}$, the analogous results are $\frac{3}{2 \sqrt{2}} 
\approx 1.06066$ and
1; for $\{6,15\}$, they are $\frac{2}{3}$ 
 and
$\frac{1}{2}$; and for $\{8,9\}$, $\frac{3 \sqrt{3}}{4}$ and 
$\frac{\sqrt{3}}{2}$,
for a separability probability of boundary states 
of $\frac{2}{3}$.
For the last $\{9,15\}$ of the five scenarios, we have $\frac{2}{\sqrt{3}}$ and 1.
Let us now present these results in the following array form (which we will 
adopt for our more extensive results further below):
\begin{equation} \label{boundary1}
\left(
\begin{array}{lllll}
 \{3,6\} & \frac{3}{2} & \frac{1}{2} & \frac{1}{3} &
   0.333333 \\
 \{6,8\} & \frac{3}{2 \sqrt{2}} & 1 & \frac{2 \sqrt{2}}{3}
   & 0.942809 \\
 \{6,15\} & \frac{2}{3} & \frac{1}{2} & \frac{3}{4} & 0.75
   \\
 \{8,9\} & \frac{3 \sqrt{3}}{4} & \frac{\sqrt{3}}{2} &
   \frac{2}{3} & 0.666667 \\
 \{9,15\} & \frac{2}{\sqrt{3}} & 1 & \frac{\sqrt{3}}{2} &
   0.866025
\end{array}
\right).
\end{equation}
\subsection{Length of Separability-Nonseparability {\it Interior} Boundary}
In the following array, we present the HS-length of the common border
separating the nonseparable (entangled) states from the separable ones.
The states lying along this interior 
border generically have {\it nondegenerate}
spectra.
\begin{equation}
\left(
\begin{array}{lllll}
 \{3,6\} & \{6,8\} & \{6,15\} & \{8,9\} & \{9,15\} \\
 \frac{1}{2} & 1 & 1 & \frac{\sqrt{3}}{4} & \frac{1}{2}
\end{array}
\right).
\end{equation}
\section{The qubit-qutrit case $n=6$} \label{secN=6}
\subsection{$3 \times2$ Decomposition} \label{n=6a}
Moving on from the $n=4$ case specifically studied by Jac\'obczyk and
 Siennicki to $n=6$ (cf. \cite{slaterPRA}), 
we compute the partial transposes of the $6 \times 6$ density matrices,
corresponding to two-dimensional sections of the set of Bloch vectors.
We, first, transpose in place the ($2^2$) 
four $3 \times 3$ blocks of the density matrices. 
By the Peres-Horodecki criterion, such density matrices with {\it positive} partial 
transposes must be {\it separable}.

We obtained the following results, presented in the same manner as (\ref{n=4case}).
\begin{equation} \label{n=6case1}
\left(
\begin{array}{lllll}
 \{1,13\} & 48 & \frac{4}{9} & \frac{\pi }{4} & 0.785398
   \\
 \{3,11\} & 4 & \frac{8 \sqrt{2}}{27} & \frac{1}{\sqrt{2}}
   & 0.707107 \\
 \{3,13\} & 4 & \frac{4}{9} & \frac{5}{6} & 0.833333 \\
 \{3,25\} & 4 & \frac{8 \sqrt{2}}{27} & \frac{5}{4
   \sqrt{2}} & 0.883883 \\
 \{8,13\} & 4 & \frac{2}{3} & \frac{1}{\sqrt{3}} &
   0.577350 \\
 \{8,25\} & 4 & \frac{\sqrt{2}}{3} & \sqrt{\frac{2}{3}} &
   0.816497 \\
 \{11,15\} & 4 & \frac{4 \sqrt{2} \pi }{27} &
   \frac{1}{3}+\frac{3 \sqrt{3}}{4 \pi } & 0.746830 \\
 \{11,24\} & 2 & \frac{25 \sqrt{\frac{5}{2}}}{72} &
   \frac{2}{5}+\frac{1}{2} \sin
   ^{-1}\left(\frac{4}{5}\right) & 0.863648 \\
 \{13,24\} & 2 & \frac{25 \sqrt{\frac{5}{2}}}{72} &
   \frac{8}{75} \left(-2+5 \sqrt{5}\right) & 0.979236 \\
 \{13,35\} & 4 & \frac{4 \sqrt{\frac{3}{5}}}{5} &
   \frac{1}{12} \left(5+3 \sqrt{5} \csc
   ^{-1}\left(\frac{3}{\sqrt{5}}\right)\right) & 0.886838
   \\
 \{15,16\} & 4 & \frac{32 \sqrt{2}}{81} & \frac{1}{32}
   \left(9 \sqrt{3}+4 \pi \right) & 0.879838 \\
 \{16,24\} & 2 & \frac{25}{144} \sqrt{\frac{5}{2}} \pi  &
   \frac{4+5 \sin ^{-1}\left(\frac{4}{5}\right)}{5 \pi } &
   0.549815 \\
 \{20,24\} & 2 & \frac{25}{144} \sqrt{\frac{5}{2}} \pi  &
   \frac{92+75 \sin ^{-1}\left(\frac{4}{5}\right)}{75 \pi
   } & 0.685627 \\
 \{24,25\} & 2 & \frac{25}{27 \sqrt{2}} & 1-\frac{2}{5
   \sqrt{5}} & 0.821115 \\
 \{24,27\} & 2 & \frac{25}{27 \sqrt{2}} & \frac{92+75 \cos
   ^{-1}\left(\frac{3}{5}\right)}{80 \sqrt{5}} & 0.903076
   \\
 \{25,35\} & 4 & \frac{\sqrt{3} \pi }{5} &
   \frac{\sqrt{5}+3 \csc
   ^{-1}\left(\frac{3}{\sqrt{5}}\right)}{3 \pi } &
   0.504975
\end{array}
\right).
\end{equation}
\subsubsection{Boundary states}
In the following array, we list, first the scenario pair, then, the HS-length
of the boundary states (those with degenerate spectra), then, the HS-length of
those boundary states which are separable, then, the separability probability
and a numerical approximation to it.
\begin{equation}
\left(
\begin{array}{lllll}
 \{1,13\} & \frac{2}{3} & 0 & 0 & 0. \\
 \{3,11\} & 1 & \frac{1}{3} & \frac{1}{3} & 0.333333 \\
 \{3,13\} & \frac{2}{3} & \frac{1}{3} & \frac{1}{2} & 0.5
   \\
 \{3,25\} & 1 & \frac{1}{3} & \frac{1}{3} & 0.333333 \\
 \{8,13\} & \frac{2}{\sqrt{3}} & \frac{1}{2 \sqrt{3}} &
   \frac{1}{4} & 0.25 \\
 \{8,25\} & \frac{\sqrt{3}}{2} & \frac{1}{\sqrt{3}} &
   \frac{2}{3} & 0.666667 \\
 \{11,15\} & \frac{4}{3 \sqrt{3}} & \frac{2}{3} &
   \frac{\sqrt{3}}{2} & 0.866025 \\
 \{11,24\} & \frac{5}{12} & \frac{1}{3} & \frac{4}{5} &
   0.8 \\
 \{13,24\} & \frac{5}{6} & \frac{\sqrt{5}}{3} &
   \frac{2}{\sqrt{5}} & 0.894427 \\
 \{13,35\} & \frac{2}{5} & \frac{1}{3} & \frac{5}{6} &
   0.833333 \\
 \{15,16\} & \frac{4 \sqrt{\frac{2}{3}}}{3} &
   \sqrt{\frac{2}{3}} & \frac{3}{4} & 0.75 \\
 \{16,24\} & \frac{5}{6} & \frac{2}{3} & \frac{4}{5} & 0.8
   \\
 \{20,24\} & \frac{5}{6} & \frac{2}{3} & \frac{4}{5} & 0.8
   \\
 \{24,25\} & \frac{5 \sqrt{\frac{5}{2}}}{6} &
   \frac{\sqrt{\frac{5}{2}}}{6} & \frac{1}{5} & 0.2 \\
 \{24,27\} & \frac{5 \sqrt{\frac{5}{2}}}{6} &
   \frac{\sqrt{10}}{3} & \frac{4}{5} & 0.8 \\
 \{25,35\} & \frac{2}{\sqrt{5}} & \frac{2}{3} &
   \frac{\sqrt{5}}{3} & 0.745356
\end{array}
\right).
\end{equation}
\subsubsection{Length of Separability-Nonseparability {\it Interior} Boundary}
In the following arrays, we present the HS-length of the common border
separating the nonseparable (entangled) states from the separable ones 
for each specific scenario.
\begin{equation}
\left(
\begin{array}{llllllll}
 \{1,13\} & \{3,11\} & \{3,13\} & \{3,25\} & \{8,13\} &
   \{8,25\} & \{11,15\} & \{11,24\} \\
 \frac{2}{3} & \frac{1}{3} & \frac{1}{3} & \frac{1}{3} &
   \frac{1}{\sqrt{3}} & \frac{1}{2 \sqrt{3}} & \frac{2}{3}
   & \frac{2}{3}
\end{array}
\right)
\end{equation}
\begin{displaymath}
\left(
\begin{array}{llllllll}
 \{13,24\} & \{13,35\} & \{15,16\} & \{16,24\} & \{20,24\}
   & \{24,25\} & \{24,27\} & \{25,35\} \\
 \frac{\sqrt{5}}{3} & \frac{2}{3} &
   \frac{\sqrt{\frac{2}{3}}}{3} & \frac{1}{3} &
   \frac{2}{3} & \frac{\sqrt{10}}{3} &
   \frac{\sqrt{\frac{5}{2}}}{6} & \frac{1}{3}
\end{array}
\right).
\end{displaymath}

\subsection{$2 \times 3$ Decomposition} \label{n=6b}
Here, we compute the partial transposes of the 
same collection of $6 \times 6$ density matrices,
corresponding to two-dimensional sections of the set of Bloch vectors,
by transposing in place the ($3^2$)
 nine  $2 \times 2$ blocks of the density matrices --- rather than the 
four ($2^2) 3 \times 3$ blocks as previously (sec.~\ref{n=6a}).

We obtained the following results. 

\begin{equation}
\left(
\begin{array}{lllll}
 \{3,6\} & 8 & \frac{8 \sqrt{2}}{27} & \frac{1}{\sqrt{2}}
   & 0.707107 \\
 \{4,18\} & 48 & \frac{4}{9} & \frac{\pi }{4} & 0.785398
   \\
 \{6,8\} & 2 & \frac{1}{8} \sqrt{\frac{3}{2}} \pi  &
   \frac{26 \sqrt{2}+27 \tan ^{-1}\left(2
   \sqrt{2}\right)}{27 \pi } & 0.825312 \\
 \{6,15\} & 2 & \frac{16 \sqrt{\frac{2}{3}}}{27} &
   \frac{1}{24} \left(9+2 \sqrt{3} \pi \right) & 0.828450
   \\
 \{8,9\} & 2 & \frac{\sqrt{2}}{3} & \frac{52+27 \sqrt{2}
   \sec ^{-1}(3)}{48 \sqrt{6}} & 0.842035 \\
 \{8,22\} & 2 & \frac{1}{\sqrt{3}} & \frac{8}{9} &
   0.888889 \\
 \{8,29\} & 2 & \frac{2}{3} & \frac{4}{3 \sqrt{3}} &
   0.769800 \\
 \{9,15\} & 2 & \frac{4 \sqrt{2} \pi }{27} &
   \frac{1}{3}+\frac{\sqrt{3}}{2 \pi } & 0.608998 \\
 \{15,22\} & 2 & \frac{32 \sqrt{\frac{2}{3}}}{27} &
   \frac{1}{2} & 0.500000 \\
 \{15,29\} & 2 & \frac{32 \sqrt{2}}{81} &
   \frac{\sqrt{3}}{2} & 0.866025 \\
 \{18,24\} & 4 & \frac{25}{144} \sqrt{\frac{5}{2}} \pi  &
   \frac{92+75 \sin ^{-1}\left(\frac{4}{5}\right)}{75 \pi
   } & 0.685627 \\
 \{18,35\} & 4 & \frac{4 \sqrt{\frac{3}{5}}}{5} &
   \frac{1}{12} \left(5+3 \sqrt{5} \csc
   ^{-1}\left(\frac{3}{\sqrt{5}}\right)\right) & 0.886838
   \\
 \{24,25\} & 4 & \frac{25}{27 \sqrt{2}} & \frac{92+75 \cos
   ^{-1}\left(\frac{3}{5}\right)}{80 \sqrt{5}} & 0.903076
   \\
 \{25,35\} & 4 & \frac{\sqrt{3} \pi }{5} &
   \frac{\sqrt{5}+3 \csc
   ^{-1}\left(\frac{3}{\sqrt{5}}\right)}{3 \pi } &
   0.504975
\end{array}
\right).
\end{equation}
There are now 
only fourteen rows, while in 
the preceding  qubit-qutrit analysis (\ref{n=6case1}) there were sixteen.
In both analyses, though,  there are 48 unordered pairs of Lie generators
which yield HS separability probabilities equal to $\frac{\pi}{4}$.

\subsubsection{Boundary states}
Here, we again present the results, restricting consideration to the
boundary (degenerate spectra) states, in the same form as previously 
(\ref{boundary1}).
\begin{equation}
\left(
\begin{array}{lllll}
 \{3,6\} & 1 & \frac{1}{3} & \frac{1}{3} & 0.333333 \\
 \{4,18\} & \frac{2}{3} & 0 & 0 & 0. \\
 \{6,8\} & \frac{1}{\sqrt{2}} & \frac{2}{3} & \frac{2
   \sqrt{2}}{3} & 0.942809 \\
 \{6,15\} & \frac{4}{9} & \frac{1}{3} & \frac{3}{4} & 0.75
   \\
 \{8,9\} & \frac{\sqrt{3}}{2} & \frac{1}{\sqrt{3}} &
   \frac{2}{3} & 0.666667 \\
 \{8,22\} & \frac{\sqrt{3}}{2} & \frac{1}{\sqrt{3}} &
   \frac{2}{3} & 0.666667 \\
 \{8,29\} & \frac{2}{\sqrt{3}} & \frac{1}{2 \sqrt{3}} &
   \frac{1}{4} & 0.25 \\
 \{9,15\} & \frac{4}{3 \sqrt{3}} & \frac{2}{3} &
   \frac{\sqrt{3}}{2} & 0.866025 \\
 \{15,22\} & \frac{5 \sqrt{\frac{2}{3}}}{3} &
   \frac{\sqrt{\frac{2}{3}}}{3} & \frac{1}{5} & 0.2 \\
 \{15,29\} & \frac{4 \sqrt{\frac{2}{3}}}{3} &
   \sqrt{\frac{2}{3}} & \frac{3}{4} & 0.75 \\
 \{18,24\} & \frac{5}{6} & \frac{2}{3} & \frac{4}{5} & 0.8
   \\
 \{18,35\} & \frac{2}{5} & \frac{1}{3} & \frac{5}{6} &
   0.833333 \\
 \{24,25\} & \frac{5 \sqrt{\frac{5}{2}}}{6} &
   \frac{\sqrt{10}}{3} & \frac{4}{5} & 0.8 \\
 \{25,35\} & \frac{2}{\sqrt{5}} & \frac{2}{3} &
   \frac{\sqrt{5}}{3} & 0.745356
\end{array}
\right)
\end{equation}
\subsubsection{Length of Separability-Nonseparability {\it Interior} Boundary}
In the following arrays, we present the HS-length of the common border
separating the nonseparable (entangled) states from the separable ones
for each specific scenario.
\begin{equation}
\left(
\begin{array}{lllllll}
 \{3,6\} & \{4,18\} & \{6,8\} & \{6,15\} & \{8,9\} &
   \{8,22\} & \{8,29\} \\
 \frac{1}{3} & \frac{2}{3} & \frac{2}{3} & \frac{2}{3} &
   \frac{1}{2 \sqrt{3}} & \frac{1}{2 \sqrt{3}} &
   \frac{1}{\sqrt{3}}
\end{array}
\right)
\end{equation}
\begin{displaymath}
\left(
\begin{array}{lllllll}
 \{9,15\} & \{15,22\} & \{15,29\} & \{18,24\} & \{18,35\}
   & \{24,25\} & \{25,35\} \\
 \frac{1}{3} & \sqrt{\frac{2}{3}} &
   \frac{\sqrt{\frac{2}{3}}}{3} & \frac{2}{3} &
   \frac{2}{3} & \frac{\sqrt{\frac{5}{2}}}{6} &
   \frac{1}{3}
\end{array}
\right).
\end{displaymath}
\subsection{Biseparable HS probabilities}
Now, we determine which of the two-dimensional set of $6 \times 6$ density matrices have positive partial transposes for {\it both} forms of partial
transposition used in sec.~\ref{n=6a} \ref{n=6b}. 
The results we obtained were:
\begin{equation}
\left(
\begin{array}{lllll}
 \{1,13\} & 88 & \frac{4}{9} & \frac{\pi }{4} & 0.785398
   \\
 \{3,6\} & 12 & \frac{8 \sqrt{2}}{27} & \frac{1}{\sqrt{2}}
   & 0.707107 \\
 \{3,13\} & 4 & \frac{4}{9} & \frac{5}{6} & 0.833333 \\
 \{3,27\} & 2 & \frac{8 \sqrt{2}}{27} & \frac{5}{4
   \sqrt{2}} & 0.883883 \\
 \{6,8\} & 2 & \frac{1}{8} \sqrt{\frac{3}{2}} \pi  &
   \frac{26 \sqrt{2}+27 \tan ^{-1}\left(2
   \sqrt{2}\right)}{27 \pi } & 0.825312 \\
 \{6,15\} & 2 & \frac{16 \sqrt{\frac{2}{3}}}{27} &
   \frac{1}{24} \left(9+2 \sqrt{3} \pi \right) & 0.828450
   \\
 \{8,9\} & 2 & \frac{\sqrt{2}}{3} & \frac{52+27 \sqrt{2}
   \sec ^{-1}(3)}{48 \sqrt{6}} & 0.842035 \\
 \{8,13\} & 4 & \frac{2}{3} & \frac{1}{\sqrt{3}} &
   0.577350 \\
\end{array}
\right).
\end{equation}
\subsubsection{Boundary States}
Concerning the corresponding one-dimensional (exterior) boundary (degenerate 
spectra) states we found:
\begin{equation}
\left(
\begin{array}{lllll}
 \{1,13\} & \frac{2}{3} & 0 & 0 & 0. \\
 \{3,6\} & 1 & \frac{1}{3} & \frac{1}{3} & 0.333333 \\
 \{3,13\} & \frac{2}{3} & \frac{1}{3} & \frac{1}{2} & 0.5
   \\
 \{3,27\} & 1 & \frac{1}{3} & \frac{1}{3} & 0.333333 \\
 \{6,8\} & \frac{1}{\sqrt{2}} & \frac{2}{3}
   & \frac{2 \sqrt{2}}{3} & 0.942809 \\
 \{6,15\} & \frac{4}{9} & \frac{1}{3} & \frac{3}{4} & 0.75
   \\
 \{8,9\} & \frac{\sqrt{3}}{2} & \frac{1}{\sqrt{3}} &
   \frac{2}{3} & 0.666667 \\
 \{8,13\} & \frac{2}{\sqrt{3}} & \frac{1}{2 \sqrt{3}} &
   \frac{1}{4} & 0.25 \\
 \{8,22\} & \frac{\sqrt{3}}{2} & \frac{1}{\sqrt{3}} &
   \frac{2}{3} & 0.666667 \\
 \{8,25\} & \frac{\sqrt{3}}{2} & \frac{1}{\sqrt{3}} &
   \frac{2}{3} & 0.666667 \\
 \{8,29\} & \frac{2}{\sqrt{3}} & \frac{1}{2 \sqrt{3}} &
   \frac{1}{4} & 0.25 \\
 \{9,15\} & \frac{4}{3 \sqrt{3}} & \frac{2}{3} &
   \frac{\sqrt{3}}{2} & 0.866025 \\
 \{11,15\} & \frac{4}{3 \sqrt{3}} & \frac{2}{3} &
   \frac{\sqrt{3}}{2} & 0.866025 \\
 \{11,24\} & \frac{5}{12} & \frac{1}{3} & \frac{4}{5} &
   0.8 \\
 \{13,24\} & \frac{5}{6} & \frac{\sqrt{5}}{3} &
   \frac{2}{\sqrt{5}} & 0.894427 \\
 \{13,35\} & \frac{2}{5} & \frac{1}{3} & \frac{5}{6} &
   0.833333 \\
 \{15,16\} & \frac{4 \sqrt{\frac{2}{3}}}{3} &
   \sqrt{\frac{2}{3}} & \frac{3}{4} & 0.75 \\
 \{15,22\} & \frac{5 \sqrt{\frac{2}{3}}}{3} &
   \frac{\sqrt{\frac{2}{3}}}{3} & \frac{1}{5} & 0.2 \\
 \{15,29\} & \frac{4 \sqrt{\frac{2}{3}}}{3} &
   \sqrt{\frac{2}{3}} & \frac{3}{4} & 0.75 \\
 \{16,24\} & \frac{5}{6} & \frac{2}{3} & \frac{4}{5} & 0.8
   \\
 \{18,24\} & \frac{5}{6} & \frac{2}{3} & \frac{4}{5} & 0.8
   \\
 \{24,25\} & \frac{5 \sqrt{\frac{5}{2}}}{6} & 0 & 0 & 0.
   \\
 \{24,27\} & \frac{5 \sqrt{\frac{5}{2}}}{6} &
   \frac{\sqrt{10}}{3} & \frac{4}{5} & 0.8 \\
 \{25,35\} & \frac{2}{\sqrt{5}} & \frac{2}{3} &
   \frac{\sqrt{5}}{3} & 0.745356
\end{array}
\right).
\end{equation}
\subsubsection{Length of Biseparability-Nonbiseparability {\it Interior} Boundary}
In the following array, we present the HS-length of the common border
separating the (generically nondegenerate) 
biseparable states from the non-biseparable ones
for each specific scenario.
\begin{equation}
\left(
\begin{array}{ll}
 \{1,13\} & \frac{2}{3} \\
 \{3,6\} & \frac{2}{3} \\
 \{3,13\} & \frac{2}{3} \\
 \{3,27\} & \frac{2}{3} \\
 \{6,8\} & \frac{4}{3} \\
 \{6,15\} & 1 \\
 \{8,9\} & \frac{\sqrt{3}}{2} \\
 \{8,13\} & \frac{\sqrt{3}}{2} \\
 \{8,22\} & \frac{\sqrt{3}}{2} \\
 \{8,25\} & \frac{\sqrt{3}}{2} \\
 \{8,29\} & \frac{\sqrt{3}}{2} \\
 \{9,15\} & 1 \\
 \{11,15\} & \frac{4}{3} \\
 \{11,24\} & 1 \\
 \{13,24\} & \frac{2 \sqrt{5}}{3} \\
 \{13,35\} & 1 \\
 \{15,16\} & \frac{4 \sqrt{\frac{2}{3}}}{3} \\
 \{15,22\} & \frac{4 \sqrt{\frac{2}{3}}}{3} \\
 \{15,29\} & \frac{4 \sqrt{\frac{2}{3}}}{3} \\
 \{16,24\} & 1 \\
 \{18,24\} & \frac{4}{3} \\
 \{24,25\} &
   \frac{\sqrt{\frac{5}{2}}}{6}+\frac{\sqrt{10}}{3} \\
 \{24,27\} &
   \frac{\sqrt{\frac{5}{2}}}{6}+\frac{\sqrt{10}}{3} \\
 \{25,35\} & \frac{2}{3}
\end{array}
\right).
\end{equation}
\section{The case $n=8$} \label{secN=8}
\subsection{$4 \times 2$ Decomposition}
Here, we compute the partial transposes of the $8 \times 8$ density matrices,
corresponding to two-dimensional sections of the set of Bloch vectors,
by, first,  transposing in place the ($2^2$)
four $4 \times 4$ blocks of the density matrices. The results are
\begin{equation} \label{eight4x2}
\left(
\begin{array}{lllll}
 \{1,20\} & 192 & \frac{1}{4} & \frac{\pi }{4} & 0.785398
   \\
 \{3,18\} & 4 & \frac{1}{3 \sqrt{2}} & \frac{1}{\sqrt{2}}
   & 0.707107 \\
 \{3,20\} & 8 & \frac{1}{4} & \frac{5}{6} & 0.833333 \\
 \{3,36\} & 8 & \frac{1}{3 \sqrt{2}} & \frac{5}{4
   \sqrt{2}} & 0.883883 \\
 \{8,20\} & 4 & \frac{3}{8} & \frac{1}{\sqrt{3}} &
   0.577350 \\
 \{8,22\} & 4 & \frac{3 \sqrt{3}}{16} & \frac{7}{9} &
   0.777778 \\
 \{8,36\} & 4 & \frac{3}{8 \sqrt{2}} & \sqrt{\frac{2}{3}}
   & 0.816497 \\
 \{8,42\} & 2 & \frac{3 \sqrt{3}}{16} & \frac{8}{9} &
   0.888889 \\
\end{array}
\right).
\end{equation}
\subsubsection{Boundary states}
The results pertaining to the {\it one}-dimensional (exterior) 
boundary (generically degenerate) states were:
\begin{equation}
\left(
\begin{array}{lllll}
 \{1,20\} & \frac{1}{2} & 0 & 0 & 0. \\
 \{3,18\} & \frac{3}{4} & \frac{1}{4} & \frac{1}{3} &
   0.333333 \\
 \{3,20\} & \frac{1}{2} & \frac{1}{4} & \frac{1}{2} & 0.5
   \\
 \{3,36\} & \frac{3}{4} & \frac{1}{4} & \frac{1}{3} &
   0.333333 \\
 \{8,20\} & \frac{\sqrt{3}}{2} & \frac{\sqrt{3}}{8} &
   \frac{1}{4} & 0.25 \\
 \{8,22\} & \frac{3 \sqrt{3}}{8} & \frac{\sqrt{3}}{8} &
   \frac{1}{3} & 0.333333 \\
 \{8,36\} & \frac{3 \sqrt{3}}{8} & \frac{\sqrt{3}}{4} &
   \frac{2}{3} & 0.666667 \\
 \{8,42\} & \frac{3 \sqrt{3}}{8} & \frac{\sqrt{3}}{4} &
   \frac{2}{3} & 0.666667 \\
\end{array}
\right).
\end{equation}
Here, we see the appearance of 
(fully entangled) domains that have no separable component, at all.
\subsection{$2 \times 4$ Decomposition}
Now, we compute the partial transposes of the $8 \times 8$ density matrices,
corresponding to two-dimensional sections of the set of Bloch vectors,
by transposing in place the ($4^4$)
sixteen  $2 \times 2$ 
blocks of the density matrices.
We obtained the following results.
\begin{equation} \label{eight2x4}
\left(
\begin{array}{lllll}
 \{3,6\} & 12 & \frac{1}{3 \sqrt{2}} & \frac{1}{\sqrt{2}}
   & 0.707107 \\
 \{4,18\} & 192 & \frac{1}{4} & \frac{\pi }{4} & 0.785398
   \\
 \{6,8\} & 2 & \frac{9}{128} \sqrt{\frac{3}{2}} \pi  &
   \frac{26 \sqrt{2}+27 \tan ^{-1}\left(2
   \sqrt{2}\right)}{27 \pi } & 0.825312 \\
 \{6,15\} & 2 & \frac{\sqrt{\frac{2}{3}}}{3} &
   \frac{1}{24} \left(9+2 \sqrt{3} \pi \right) & 0.828450
   \\
 \{8,9\} & 2 & \frac{3}{8 \sqrt{2}} & \frac{52+27 \sqrt{2}
   \sec ^{-1}(3)}{48 \sqrt{6}} & 0.842035 \\
 \{8,22\} & 4 & \frac{3 \sqrt{3}}{16} & \frac{8}{9} &
   0.888889 \\
 \{8,29\} & 4 & \frac{3}{8} & \frac{4}{3 \sqrt{3}} &
   0.769800 \\
 \{9,15\} & 2 & \frac{\pi }{6 \sqrt{2}} &
   \frac{1}{3}+\frac{\sqrt{3}}{2 \pi } & 0.608998 \\
\end{array}
\right).
\end{equation}
We see that there are fewer rows in (\ref{eight2x4}) than in (\ref{eight4x2}), 
obtained by the alternative form of partial transposition.
\subsubsection{Boundary states}
Our analysis of the HS-{\it lengths} of the corresponding boundary states yielded:
\begin{equation}
\left(
\begin{array}{lllll}
 \{3,6\} & \frac{3}{4} & \frac{1}{4} & \frac{1}{3} &
   0.333333 \\
 \{4,18\} & \frac{1}{2} & 0 & 0 & 0. \\
 \{6,8\} & \frac{3}{4 \sqrt{2}} & \frac{1}{2} & \frac{2
   \sqrt{2}}{3} & 0.942809 \\
 \{6,15\} & \frac{1}{3} & \frac{1}{4} & \frac{3}{4} & 0.75
   \\
 \{8,9\} & \frac{3 \sqrt{3}}{8} & \frac{\sqrt{3}}{4}
  & \frac{2}{3} & 0.666667 \\
 \{8,22\} & \frac{3 \sqrt{3}}{8} & \frac{\sqrt{3}}{4} &
   \frac{2}{3} & 0.666667 \\
 \{8,29\} & \frac{\sqrt{3}}{2} & \frac{\sqrt{3}}{8} &
   \frac{1}{4} & 0.25 \\
 \{9,15\} & \frac{1}{\sqrt{3}} & \frac{1}{2} &
   \frac{\sqrt{3}}{2} & 0.866025 \\
 \{15,22\} & \frac{5}{2 \sqrt{6}} & \frac{1}{2 \sqrt{6}} &
   \frac{1}{5} & 0.2 \\
 \{15,29\} & \sqrt{\frac{2}{3}} &
   \frac{\sqrt{\frac{3}{2}}}{2} & \frac{3}{4} & 0.75 \\
 \{18,24\} & \frac{5}{8} & \frac{1}{2} & \frac{4}{5} & 0.8
   \\
 \{18,35\} & \frac{3}{10} & \frac{1}{4} & \frac{5}{6} &
   0.833333 \\
 \{24,25\} & \frac{5 \sqrt{\frac{5}{2}}}{8} &
   \frac{\sqrt{\frac{5}{2}}}{2} & \frac{4}{5} & 0.8 \\
 \{24,46\} & \frac{5 \sqrt{\frac{5}{2}}}{8} &
   \frac{\sqrt{\frac{5}{2}}}{2} & \frac{4}{5} & 0.8 \\
 \{24,57\} & \frac{3 \sqrt{\frac{5}{2}}}{4} &
   \frac{\sqrt{\frac{5}{2}}}{8} & \frac{1}{6} & 0.166667
   \\
 \{25,35\} & \frac{3}{2 \sqrt{5}} & \frac{1}{2} &
   \frac{\sqrt{5}}{3} & 0.745356 \\
 \{35,46\} & \frac{7 \sqrt{\frac{3}{5}}}{4} &
   \frac{\sqrt{\frac{3}{5}}}{4} & \frac{1}{7} & 0.142857
   \\
 \{35,57\} & \frac{3 \sqrt{\frac{3}{5}}}{2} &
   \frac{\sqrt{15}}{4} & \frac{5}{6} & 0.833333 \\
 \{38,48\} & \frac{7}{4 \sqrt{6}} & \frac{1}{2} & \frac{2
   \sqrt{6}}{7} & 0.699854 \\
 \{38,63\} & \frac{2}{7} & \frac{1}{4} & \frac{7}{8} &
   0.875 \\
 \{48,49\} & \frac{7 \sqrt{\frac{7}{3}}}{8} &
   \frac{\sqrt{21}}{4} & \frac{6}{7} & 0.857143 \\
 \{49,63\} & \frac{2}{\sqrt{7}} & \frac{1}{2} &
   \frac{\sqrt{7}}{4} & 0.661438
\end{array}
\right).
\end{equation}
\subsection{Bi-PPT} \label{secBi}
Here, we obtain the probabilities that an $8 \times 8$ density matrix
will have a positive partial transpose, under {\it both} forms
of partial transposition employed immediately above.
Our results were:

\begin{equation}
\left(
\begin{array}{lllll}
 \{1,20\} & 288 & \frac{1}{4} & \frac{\pi }{4} & 0.785398
   \\
 \{3,6\} & 12 & \frac{1}{3 \sqrt{2}} & \frac{1}{\sqrt{2}}
   & 0.707107 \\
 \{3,20\} & 8 & \frac{1}{4} & \frac{5}{6} & 0.833333 \\
 \{3,36\} & 4 & \frac{1}{3 \sqrt{2}} & \frac{5}{4
   \sqrt{2}} & 0.883883 \\
 \{6,8\} & 2 & \frac{9}{128} \sqrt{\frac{3}{2}} \pi  &
   \frac{26 \sqrt{2}+27 \tan ^{-1}\left(2
   \sqrt{2}\right)}{27 \pi } & 0.825312 \\
 \{6,15\} & 2 & \frac{\sqrt{\frac{2}{3}}}{3} &
   \frac{1}{24} \left(9+2 \sqrt{3} \pi \right) & 0.828450
   \\
 \{8,9\} & 2 & \frac{3}{8 \sqrt{2}} & \frac{52+27 \sqrt{2}
   \sec ^{-1}(3)}{48 \sqrt{6}} & 0.842035 \\
 \{8,20\} & 4 & \frac{3}{8} & \frac{1}{\sqrt{3}} &
   0.577350 \\
 \{8,22\} & 2 & \frac{3 \sqrt{3}}{16} & \frac{2}{3} &
   0.666667 \\
 \{8,31\} & 2 & \frac{3 \sqrt{3}}{16} & \frac{7}{9} &
   0.777778 \\
 \{8,36\} & 4 & \frac{3}{8 \sqrt{2}} & \sqrt{\frac{2}{3}}
   & 0.816497 \\
 \{8,42\} & 2 & \frac{3 \sqrt{3}}{16} & \frac{8}{9} &
   0.888889 \\
 \{8,49\} & 4 & \frac{3}{8 \sqrt{2}} & \frac{7}{3
   \sqrt{6}} & 0.952579 \\
 \{8,53\} & 2 & \frac{3}{8} & \frac{4}{3 \sqrt{3}} &
   0.769800 \\
 \{9,15\} & 2 & \frac{\pi }{6 \sqrt{2}} &
   \frac{1}{3}+\frac{\sqrt{3}}{2 \pi } & 0.608998 \\
 \{15,22\} & 6 & \frac{2 \sqrt{\frac{2}{3}}}{3} &
   \frac{1}{2} & 0.500000 \\
 \{15,29\} & 8 & \frac{2 \sqrt{2}}{9} & \frac{\sqrt{3}}{2}
   & 0.866025 \\
 \{18,24\} & 6 & \frac{25}{256} \sqrt{\frac{5}{2}} \pi  &
   \frac{92+75 \sin ^{-1}\left(\frac{4}{5}\right)}{75 \pi
   } & 0.685627 \\
 \{18,35\} & 4 & \frac{9 \sqrt{\frac{3}{5}}}{20} &
   \frac{1}{12} \left(5+3 \sqrt{5} \csc
   ^{-1}\left(\frac{3}{\sqrt{5}}\right)\right) & 0.886838
\end{array}
\right)
\end{equation}
\begin{displaymath}
\left(
\begin{array}{lllll}
 \{20,35\} & 2 & \frac{9 \sqrt{\frac{3}{5}}}{20} &
   \frac{1}{108} \left(-25+24 \sqrt{30}\right) & 0.985680
   \\
 \{20,48\} & 4 & \frac{49 \sqrt{\frac{7}{3}}}{192} &
   \frac{1}{28} \left(12+7 \sqrt{6} \sin
   ^{-1}\left(\frac{2 \sqrt{6}}{7}\right)\right) &
   0.903278 \\
 \{22,48\} & 4 & \frac{49 \sqrt{\frac{7}{3}}}{192} &
   \frac{4}{147} \left(-9+7 \sqrt{42}\right) & 0.989529 \\
 \{22,63\} & 10 & \frac{8}{7 \sqrt{7}} & \frac{1}{16}
   \left(7+4 \sqrt{7} \csc
   ^{-1}\left(\frac{4}{\sqrt{7}}\right)\right) & 0.915544
   \\
 \{24,25\} & 8 & \frac{25}{48 \sqrt{2}} & \frac{92+75 \cos
   ^{-1}\left(\frac{3}{5}\right)}{80 \sqrt{5}} & 0.903076
   \\
 \{24,46\} & 2 & \frac{5 \sqrt{\frac{5}{2}}}{16} &
   \frac{14}{15} & 0.933333 \\
 \{24,57\} & 2 & \frac{25}{24 \sqrt{2}} & \frac{7}{5
   \sqrt{5}} & 0.626099 \\
 \{25,35\} & 4 & \frac{9 \sqrt{3} \pi }{80} &
   \frac{\sqrt{5}+3 \csc
   ^{-1}\left(\frac{3}{\sqrt{5}}\right)}{3 \pi } &
   0.504975 \\
\end{array}
\right).
\end{displaymath}
\subsubsection{Boundary states}
The lengths, separable lengths and separability probabilities of
the corresponding (exterior/degenerate spectra) 
boundary states are given in the following array:
\begin{equation}
\left(
\begin{array}{lllll}
 \{1,20\} & \frac{1}{2} & 0 & 0 & 0. \\
 \{3,6\} & \frac{3}{4} & \frac{1}{4} & \frac{1}{3} &
   0.333333 \\
 \{3,20\} & \frac{1}{2} & \frac{1}{4} & \frac{1}{2} & 0.5
   \\
 \{3,36\} & \frac{3}{4} & \frac{1}{4} & \frac{1}{3} &
   0.333333 \\
 \{6,8\} & \frac{3}{4 \sqrt{2}} & \frac{1}{2} & \frac{2
   \sqrt{2}}{3} & 0.942809 \\
 \{6,15\} & \frac{1}{3} & \frac{1}{4} & \frac{3}{4} & 0.75
   \\
 \{8,9\} & \frac{3 \sqrt{3}}{8} & \frac{\sqrt{3}}{4} &
   \frac{2}{3} & 0.666667 \\
 \{8,20\} & \frac{\sqrt{3}}{2} & \frac{\sqrt{3}}{8} &
   \frac{1}{4} & 0.25 \\
 \{8,22\} & \frac{3 \sqrt{3}}{8} & 0 & 0 & 0. \\
 \{8,31\} & \frac{3 \sqrt{3}}{8} & \frac{\sqrt{3}}{8} &
   \frac{1}{3} & 0.333333 \\
 \{8,36\} & \frac{3 \sqrt{3}}{8} & \frac{\sqrt{3}}{4} &
   \frac{2}{3} & 0.666667 \\
 \{8,42\} & \frac{3 \sqrt{3}}{8} & \frac{\sqrt{3}}{4} &
   \frac{2}{3} & 0.666667 \\
 \{8,49\} & \frac{3 \sqrt{3}}{8} & \frac{\sqrt{3}}{4} &
   \frac{2}{3} & 0.666667 \\
 \{8,53\} & \frac{\sqrt{3}}{2} & \frac{\sqrt{3}}{8} &
   \frac{1}{4} & 0.25 \\
 \{9,15\} & \frac{1}{\sqrt{3}} & \frac{1}{2} &
   \frac{\sqrt{3}}{2} & 0.866025 \\
 \{15,22\} & \frac{5}{2 \sqrt{6}} & \frac{1}{2 \sqrt{6}} &
   \frac{1}{5} & 0.2 \\
 \{15,29\} & \sqrt{\frac{2}{3}} &
   \frac{\sqrt{\frac{3}{2}}}{2} & \frac{3}{4} & 0.75 \\
 \{18,24\} & \frac{5}{8} & \frac{1}{2} & \frac{4}{5} & 0.8
   \\
 \{18,35\} & \frac{3}{10} & \frac{1}{4} & \frac{5}{6} &
   0.833333
\end{array}
\right)
\end{equation}
\begin{displaymath}
\left(
\begin{array}{lllll}
 \{20,35\} & \frac{3}{5} & \sqrt{\frac{3}{10}} &
   \sqrt{\frac{5}{6}} & 0.912871 \\
 \{20,48\} & \frac{7}{24} & \frac{1}{4} & \frac{6}{7} &
   0.857143 \\
 \{22,48\} & \frac{7}{12} & \frac{\sqrt{\frac{7}{6}}}{2} &
   \sqrt{\frac{6}{7}} & 0.92582 \\
 \{22,63\} & \frac{2}{7} & \frac{1}{4} & \frac{7}{8} &
   0.875 \\
 \{24,25\} & \frac{5 \sqrt{\frac{5}{2}}}{8} &
   \frac{\sqrt{\frac{5}{2}}}{2} & \frac{4}{5} & 0.8 \\
 \{24,46\} & \frac{5 \sqrt{\frac{5}{2}}}{8} &
   \frac{\sqrt{\frac{5}{2}}}{2} & \frac{4}{5} & 0.8 \\
 \{24,57\} & \frac{3 \sqrt{\frac{5}{2}}}{4} &
   \frac{\sqrt{\frac{5}{2}}}{8} & \frac{1}{6} & 0.166667
   \\
 \{25,35\} & \frac{3}{2 \sqrt{5}} & \frac{1}{2} &
   \frac{\sqrt{5}}{3} & 0.745356 \\
 \{31,35\} & \frac{3}{2 \sqrt{5}} & \frac{1}{2} &
   \frac{\sqrt{5}}{3} & 0.745356 \\
 \{35,36\} & \frac{3 \sqrt{\frac{3}{5}}}{2} &
   \frac{\sqrt{\frac{3}{5}}}{4} & \frac{1}{6} & 0.166667
   \\
 \{35,38\} & \frac{3 \sqrt{\frac{3}{5}}}{2} &
   \frac{\sqrt{15}}{4} & \frac{5}{6} & 0.833333 \\
 \{35,46\} & \frac{7 \sqrt{\frac{3}{5}}}{4} &
   \frac{\sqrt{\frac{3}{5}}}{4} & \frac{1}{7} & 0.142857
   \\
 \{35,57\} & \frac{3 \sqrt{\frac{3}{5}}}{2} &
   \frac{\sqrt{15}}{4} & \frac{5}{6} & 0.833333 \\
 \{36,48\} & \frac{7}{4 \sqrt{6}} & \frac{1}{2} & \frac{2
   \sqrt{6}}{7} & 0.699854 \\
 \{42,48\} & \frac{7}{4 \sqrt{6}} & \frac{1}{2} & \frac{2
   \sqrt{6}}{7} & 0.699854 \\
 \{48,49\} & \frac{7 \sqrt{\frac{7}{3}}}{8} & 0 & 0 & 0.
   \\
 \{48,51\} & \frac{7 \sqrt{\frac{7}{3}}}{8} &
   \frac{\sqrt{\frac{7}{3}}}{8} & \frac{1}{7} & 0.142857
   \\
 \{48,53\} & \frac{7 \sqrt{\frac{7}{3}}}{8} &
   \frac{\sqrt{21}}{4} & \frac{6}{7} & 0.857143 \\
 \{49,63\} & \frac{2}{\sqrt{7}} & \frac{1}{2} &
   \frac{\sqrt{7}}{4} & 0.661438
\end{array}
\right).
\end{displaymath}
\subsection{Tri-ppt}
Now, we derive the probabilities that an $8 \times 8$ density matrix
will have a positive partial transpose, not only 
under {\it both} forms
of partial transposition previously employed, as in sec.~\ref{secBi}, 
but also under a {\it third}
(independent) form obtained, first, applying a  
certain $8 \times 8$ {\it permutation}
matrix (\cite[eq. (3)]{zhong}) to the 
original $8 \times
8$ density matrix,  {\it then} transposing in place 
the resultant four $4 \times 4$ blocks.
We obtained the following results.

\begin{equation} \label{triseparabilitycase}
\left(
\begin{array}{lllll}
 \{1,20\} & 288 & \frac{1}{4} & \frac{\pi }{4} & 0.785398
   \\
 \{3,6\} & 12 & \frac{1}{3 \sqrt{2}} & \frac{1}{\sqrt{2}}
   & 0.707107 \\
 \{3,20\} & 4 & \frac{1}{4} & \frac{5}{6} & 0.833333 \\
 \{3,22\} & 4 & \frac{1}{4} & \frac{2}{3} & 0.666667 \\
 \{3,36\} & 4 & \frac{1}{3 \sqrt{2}} & \frac{5}{4
   \sqrt{2}} & 0.883883 \\
 \{6,8\} & 2 & \frac{9}{128} \sqrt{\frac{3}{2}} \pi  &
   \frac{26 \sqrt{2}+27 \tan ^{-1}\left(2
   \sqrt{2}\right)}{27 \pi } & 0.825312 \\
 \{6,15\} & 2 & \frac{\sqrt{\frac{2}{3}}}{3} &
   \frac{1}{24} \left(9+2 \sqrt{3} \pi \right) & 0.828450
   \\
 \{8,9\} & 2 & \frac{3}{8 \sqrt{2}} & \frac{52+27 \sqrt{2}
   \sec ^{-1}(3)}{48 \sqrt{6}} & 0.842035 \\
 \{8,20\} & 4 & \frac{3}{8} & \frac{1}{\sqrt{3}} &
   0.577350 \\
 \{8,22\} & 2 & \frac{3 \sqrt{3}}{16} & \frac{2}{3} &
   0.666667 \\
 \{8,31\} & 2 & \frac{3 \sqrt{3}}{16} & \frac{7}{9} &
   0.777778 \\
 \{8,36\} & 6 & \frac{3}{8 \sqrt{2}} & \sqrt{\frac{2}{3}}
   & 0.816497 \\
 \{8,42\} & 2 & \frac{3 \sqrt{3}}{16} & \frac{8}{9} &
   0.888889 \\
 \{8,51\} & 2 & \frac{3}{8 \sqrt{2}} & \frac{7}{3
   \sqrt{6}} & 0.952579 \\
 \{8,53\} & 2 & \frac{3}{8} & \frac{4}{3 \sqrt{3}} &
   0.769800 \\
 \{9,15\} & 2 & \frac{\pi }{6 \sqrt{2}} &
   \frac{1}{3}+\frac{\sqrt{3}}{2 \pi } & 0.608998 \\
 \{15,22\} & 6 & \frac{2 \sqrt{\frac{2}{3}}}{3} &
   \frac{1}{2} & 0.500000 \\
 \{15,29\} & 10 & \frac{2 \sqrt{2}}{9} &
   \frac{\sqrt{3}}{2} & 0.866025 \\
 \{18,24\} & 6 & \frac{25}{256} \sqrt{\frac{5}{2}} \pi  &
   \frac{92+75 \sin ^{-1}\left(\frac{4}{5}\right)}{75 \pi
   } & 0.685627 \\
 \{18,35\} & 4 & \frac{9 \sqrt{\frac{3}{5}}}{20} &
   \frac{1}{12} \left(5+3 \sqrt{5} \csc
   ^{-1}\left(\frac{3}{\sqrt{5}}\right)\right) & 0.886838
\end{array}
\right)
\end{equation}
\begin{displaymath}
\left(
\begin{array}{lllll}
 \{20,35\} & 2 & \frac{9 \sqrt{\frac{3}{5}}}{20} &
   \frac{1}{108} \left(-25+24 \sqrt{30}\right) & 0.985680
   \\
 \{20,48\} & 6 & \frac{49 \sqrt{\frac{7}{3}}}{192} &
   \frac{1}{28} \left(12+7 \sqrt{6} \sin
   ^{-1}\left(\frac{2 \sqrt{6}}{7}\right)\right) &
   0.903278 \\
 \{22,63\} & 12 & \frac{8}{7 \sqrt{7}} & \frac{1}{16}
   \left(7+4 \sqrt{7} \csc
   ^{-1}\left(\frac{4}{\sqrt{7}}\right)\right) & 0.915544
   \\
 \{24,25\} & 10 & \frac{25}{48 \sqrt{2}} & \frac{92+75
   \cos ^{-1}\left(\frac{3}{5}\right)}{80 \sqrt{5}} &
   0.903076 \\
 \{24,46\} & 2 & \frac{5 \sqrt{\frac{5}{2}}}{16} &
   \frac{14}{15} & 0.933333 \\
 \{24,57\} & 2 & \frac{25}{24 \sqrt{2}} & \frac{7}{5
   \sqrt{5}} & 0.626099 \\
 \{25,35\} & 4 & \frac{9 \sqrt{3} \pi }{80} &
   \frac{\sqrt{5}+3 \csc
   ^{-1}\left(\frac{3}{\sqrt{5}}\right)}{3 \pi } &
   0.504975 \\
 \{31,35\} & 2 & \frac{9 \sqrt{3} \pi }{80} & \frac{14
   \sqrt{5}+27 \csc
   ^{-1}\left(\frac{3}{\sqrt{5}}\right)}{27 \pi } &
   0.636783 \\
\end{array}
\right).
\end{displaymath}
Here there are only four generator pairs yielding the probability 
$\frac{5}{6}$, while there were eight in simply the ``Bi-PPT'' case
(sec.~\ref{secBi}).
\subsubsection{Boundary States}
Now, we obtained from the analysis of the one-dimensional (exterior) 
boundary states the
results:
\begin{equation}
\left(
\begin{array}{lllll}
 \{1,20\} & \frac{1}{2} & 0 & 0 & 0. \\
 \{3,6\} & \frac{3}{4} & \frac{1}{4} & \frac{1}{3} &
   0.333333 \\
 \{3,20\} & \frac{1}{2} & \frac{1}{4} & \frac{1}{2} & 0.5
   \\
 \{3,22\} & \frac{1}{2} & 0 & 0 & 0. \\
 \{3,36\} & \frac{3}{4} & \frac{1}{4} & \frac{1}{3} &
   0.333333 \\
 \{6,8\} & \frac{3}{4 \sqrt{2}} & \frac{1}{2} & \frac{2
   \sqrt{2}}{3} & 0.942809 \\
 \{6,15\} & \frac{1}{3} & \frac{1}{4} & \frac{3}{4} & 0.75
   \\
 \{8,9\} & \frac{3 \sqrt{3}}{8} & \frac{\sqrt{3}}{4} &
   \frac{2}{3} & 0.666667 \\
 \{8,20\} & \frac{\sqrt{3}}{2} & \frac{\sqrt{3}}{8} &
   \frac{1}{4} & 0.25 \\
 \{8,22\} & \frac{3 \sqrt{3}}{8} & 0 & 0 & 0. \\
 \{8,31\} & \frac{3 \sqrt{3}}{8} & \frac{\sqrt{3}}{8} &
   \frac{1}{3} & 0.333333 \\
 \{8,36\} & \frac{3 \sqrt{3}}{8} & \frac{\sqrt{3}}{4} &
   \frac{2}{3} & 0.666667 \\
 \{8,42\} & \frac{3 \sqrt{3}}{8} & \frac{\sqrt{3}}{4} &
   \frac{2}{3} & 0.666667 \\
 \{8,51\} & \frac{3 \sqrt{3}}{8} & \frac{\sqrt{3}}{4} &
   \frac{2}{3} & 0.666667 \\
 \{8,53\} & \frac{\sqrt{3}}{2} & \frac{\sqrt{3}}{8} &
   \frac{1}{4} & 0.25 \\
 \{9,15\} & \frac{1}{\sqrt{3}} & \frac{1}{2} &
   \frac{\sqrt{3}}{2} & 0.866025 \\
 \{15,22\} & \frac{5}{2 \sqrt{6}} & \frac{1}{2 \sqrt{6}} &
   \frac{1}{5} & 0.2 \\
 \{15,29\} & \sqrt{\frac{2}{3}} &
   \frac{\sqrt{\frac{3}{2}}}{2} & \frac{3}{4} & 0.75 \\
 \{18,24\} & \frac{5}{8} & \frac{1}{2} & \frac{4}{5} & 0.8
   \\
 \{18,35\} & \frac{3}{10} & \frac{1}{4} & \frac{5}{6} &
   0.833333
\end{array}
\right)
\end{equation}
\begin{displaymath}
\left(
\begin{array}{lllll}
 \{20,35\} & \frac{3}{5} & \sqrt{\frac{3}{10}} &
   \sqrt{\frac{5}{6}} & 0.912871 \\
 \{20,48\} & \frac{7}{24} & \frac{1}{4} & \frac{6}{7} &
   0.857143 \\
 \{22,63\} & \frac{2}{7} & \frac{1}{4} & \frac{7}{8} &
   0.875 \\
 \{24,25\} & \frac{5 \sqrt{\frac{5}{2}}}{8} &
   \frac{\sqrt{\frac{5}{2}}}{2} & \frac{4}{5} & 0.8 \\
 \{24,46\} & \frac{5 \sqrt{\frac{5}{2}}}{8} &
   \frac{\sqrt{\frac{5}{2}}}{2} & \frac{4}{5} & 0.8 \\
 \{24,57\} & \frac{3 \sqrt{\frac{5}{2}}}{4} &
   \frac{\sqrt{\frac{5}{2}}}{8} & \frac{1}{6} & 0.166667
   \\
 \{25,35\} & \frac{3}{2 \sqrt{5}} & \frac{1}{2} &
   \frac{\sqrt{5}}{3} & 0.745356 \\
 \{31,35\} & \frac{3}{2 \sqrt{5}} & \frac{1}{2} &
   \frac{\sqrt{5}}{3} & 0.745356 \\
 \{31,48\} & \frac{7}{12} & \frac{\sqrt{\frac{7}{6}}}{2} &
   \sqrt{\frac{6}{7}} & 0.92582 \\
 \{35,36\} & \frac{3 \sqrt{\frac{3}{5}}}{2} &
   \frac{\sqrt{\frac{3}{5}}}{4} & \frac{1}{6} & 0.166667
   \\
 \{35,38\} & \frac{3 \sqrt{\frac{3}{5}}}{2} & 0 & 0 & 0.
   \\
 \{35,46\} & \frac{7 \sqrt{\frac{3}{5}}}{4} &
   \frac{\sqrt{\frac{3}{5}}}{4} & \frac{1}{7} & 0.142857
   \\
 \{35,51\} & \frac{3 \sqrt{\frac{3}{5}}}{2} &
   \frac{\sqrt{15}}{4} & \frac{5}{6} & 0.833333 \\
 \{35,57\} & \frac{3 \sqrt{\frac{3}{5}}}{2} &
   \frac{\sqrt{15}}{4} & \frac{5}{6} & 0.833333 \\
 \{36,48\} & \frac{7}{4 \sqrt{6}} & \frac{1}{2} & \frac{2
   \sqrt{6}}{7} & 0.699854 \\
 \{42,48\} & \frac{7}{4 \sqrt{6}} & \frac{1}{2} & \frac{2
   \sqrt{6}}{7} & 0.699854 \\
 \{48,49\} & \frac{7 \sqrt{\frac{7}{3}}}{8} & 0 & 0 & 0.
   \\
 \{48,51\} & \frac{7 \sqrt{\frac{7}{3}}}{8} &
   \frac{\sqrt{\frac{7}{3}}}{8} & \frac{1}{7} & 0.142857
   \\
 \{48,53\} & \frac{7 \sqrt{\frac{7}{3}}}{8} &
   \frac{\sqrt{21}}{4} & \frac{6}{7} & 0.857143 \\
 \{49,63\} & \frac{2}{\sqrt{7}} & \frac{1}{2} &
   \frac{\sqrt{7}}{4} & 0.661438
\end{array}
\right).
\end{displaymath}
\subsubsection{Length of Triseparability-Nontriseparability {\it Interior} 
Boundary}
In the following array, we present the HS-length of the common (interior) 
border
separating the triseparable states from the non-triseparable ones
for each specific scenario.
\begin{equation}
\left(
\begin{array}{ll}
 \{1,20\} & 1 \\
 \{3,6\} & \frac{3}{4} \\
 \{3,20\} & \frac{3}{4} \\
 \{3,22\} & \frac{1}{2} \\
 \{3,36\} & \frac{3}{4} \\
 \{6,8\} & \frac{3}{2} \\
 \{6,15\} & \frac{5}{4} \\
 \{8,9\} & \frac{\sqrt{3}}{2} \\
 \{8,20\} & \frac{5 \sqrt{3}}{8} \\
 \{8,22\} & \frac{5 \sqrt{3}}{8} \\
 \{8,31\} & \frac{5 \sqrt{3}}{8} \\
 \{8,36\} & \frac{\sqrt{3}}{2} \\
 \{8,42\} & \frac{\sqrt{3}}{2} \\
 \{8,51\} & \frac{\sqrt{3}}{2} \\
 \{8,53\} & \frac{5 \sqrt{3}}{8} \\
 \{9,15\} & 1 \\
 \{15,22\} & \frac{3 \sqrt{\frac{3}{2}}}{2} \\
 \{15,29\} & \frac{7}{2 \sqrt{6}} \\
 \{18,24\} & \frac{3}{2} \\
 \{18,35\} & \frac{5}{4}
\end{array}
\right)
\end{equation}
\begin{displaymath}
\left(
\begin{array}{ll}
 \{20,35\} & 3 \sqrt{\frac{3}{10}} \\
 \{20,48\} & \frac{5}{4} \\
 \{22,63\} & 1 \\
 \{24,25\} & \frac{3 \sqrt{\frac{5}{2}}}{4} \\
 \{24,46\} & \frac{3 \sqrt{\frac{5}{2}}}{4} \\
 \{24,57\} & \frac{9 \sqrt{\frac{5}{2}}}{8} \\
 \{25,35\} & 1 \\
 \{31,35\} & \frac{3}{2} \\
 \{31,48\} & \frac{\sqrt{\frac{21}{2}}}{2} \\
 \{35,36\} & \frac{11 \sqrt{\frac{3}{5}}}{4} \\
 \{35,38\} & \frac{3 \sqrt{\frac{3}{5}}}{2} \\
 \{35,46\} & \frac{11 \sqrt{\frac{3}{5}}}{4} \\
 \{35,51\} & \frac{7 \sqrt{\frac{3}{5}}}{4} \\
 \{35,57\} & \frac{7 \sqrt{\frac{3}{5}}}{4} \\
 \{36,48\} & 1 \\
 \{42,48\} & \frac{3}{2} \\
 \{48,49\} & \frac{13 \sqrt{\frac{7}{3}}}{8} \\
 \{48,51\} & \frac{13 \sqrt{\frac{7}{3}}}{8} \\
 \{48,53\} & \sqrt{\frac{7}{3}} \\
 \{49,63\} & \frac{3}{4}
\end{array}
\right).
\end{displaymath}

\section{The qutrit-qutrit case $n=9$} \label{secN=9}
Here we only have --- since $9=3^2$ --- {\it one} option available for computing the partial transpose, 
that is transposing in place the nine $3 \times 3$ blocks of the $9 \times 9$
density matrices. We obtained the results: 
\begin{equation} \label{n=9case}
\left(
\begin{array}{lllll}
 \{1,13\} & 360 & \frac{16}{81} & \frac{\pi }{4} &
   0.785398 \\
 \{3,11\} & 8 & \frac{32 \sqrt{2}}{243} &
   \frac{1}{\sqrt{2}} & 0.707107 \\
 \{3,13\} & 8 & \frac{16}{81} & \frac{5}{6} & 0.833333 \\
 \{3,25\} & 8 & \frac{32 \sqrt{2}}{243} & \frac{5}{4
   \sqrt{2}} & 0.883883 \\
 \{8,13\} & 8 & \frac{8}{27} & \frac{1}{\sqrt{3}} &
   0.577350 \\
 \{8,25\} & 8 & \frac{4 \sqrt{2}}{27} & \sqrt{\frac{2}{3}}
   & 0.816497 \\
 \{11,15\} & 4 & \frac{16 \sqrt{2} \pi }{243} &
   \frac{1}{3}+\frac{3 \sqrt{3}}{4 \pi } & 0.746830 \\
 \{11,24\} & 2 & \frac{25 \sqrt{\frac{5}{2}}}{162} &
   \frac{2}{5}+\frac{1}{2} \sin
   ^{-1}\left(\frac{4}{5}\right) & 0.863648 \\
 \{13,24\} & 2 & \frac{25 \sqrt{\frac{5}{2}}}{162} &
   \frac{8}{75} \left(-2+5 \sqrt{5}\right) & 0.979236 \\
 \{13,35\} & 4 & \frac{16}{15 \sqrt{15}} & \frac{1}{12}
   \left(5+3 \sqrt{5} \csc
   ^{-1}\left(\frac{3}{\sqrt{5}}\right)\right) & 0.886838
   \\
 \{15,16\} & 4 & \frac{128 \sqrt{2}}{729} & \frac{1}{32}
   \left(9 \sqrt{3}+4 \pi \right) & 0.879838 \\
 \{15,44\} & 4 & \frac{32 \sqrt{\frac{2}{3}}}{81} &
   \frac{11}{12} & 0.916667 \\
 \{15,55\} & 4 & \frac{128 \sqrt{\frac{2}{3}}}{243} &
   \frac{11}{16} & 0.687500 \\
 \{16,24\} & 2 & \frac{25}{324} \sqrt{\frac{5}{2}} \pi  &
   \frac{4+5 \sin ^{-1}\left(\frac{4}{5}\right)}{5 \pi } &
   0.549815 \\
 \{20,24\} & 2 & \frac{25}{324} \sqrt{\frac{5}{2}} \pi  &
   \frac{92+75 \sin ^{-1}\left(\frac{4}{5}\right)}{75 \pi
   } & 0.685627 \\
 \{24,25\} & 2 & \frac{50 \sqrt{2}}{243} & 1-\frac{2}{5
   \sqrt{5}} & 0.821115 \\
 \{24,27\} & 2 & \frac{50 \sqrt{2}}{243} & \frac{92+75
   \cos ^{-1}\left(\frac{3}{5}\right)}{80 \sqrt{5}} &
   0.903076
\end{array}
\right)
\end{equation}
\begin{displaymath}
\left(
\begin{array}{lllll}
 \{24,44\} & 2 & \frac{100 \sqrt{2}}{243} &
   \frac{1}{\sqrt{5}} & 0.447214 \\
 \{24,46\} & 2 & \frac{10 \sqrt{10}}{81} & \frac{11}{15} &
   0.733333 \\
 \{24,55\} & 2 & \frac{50 \sqrt{2}}{243} &
   \frac{2}{\sqrt{5}} & 0.894427 \\
 \{24,59\} & 2 & \frac{10 \sqrt{10}}{81} & \frac{14}{15} &
   0.933333 \\
 \{24,70\} & 2 & \frac{50 \sqrt{2}}{243} & \frac{11}{5
   \sqrt{5}} & 0.983870 \\
 \{24,72\} & 2 & \frac{100 \sqrt{2}}{243} & \frac{7}{5
   \sqrt{5}} & 0.626099 \\
 \{25,35\} & 4 & \frac{4 \pi }{15 \sqrt{3}} &
   \frac{\sqrt{5}+3 \csc
   ^{-1}\left(\frac{3}{\sqrt{5}}\right)}{3 \pi } &
   0.504975 \\
 \{35,46\} & 4 & \frac{32 \sqrt{\frac{2}{5}}}{27} &
   \frac{1}{\sqrt{6}} & 0.408248 \\
 \{35,70\} & 4 & \frac{32 \sqrt{2}}{135} &
   \sqrt{\frac{5}{6}} & 0.912871 \\
 \{38,48\} & 8 & \frac{49}{486} \sqrt{\frac{7}{2}} \pi  &
   \frac{22 \sqrt{6}+49 \sin ^{-1}\left(\frac{2
   \sqrt{6}}{7}\right)}{49 \pi } & 0.596820 \\
 \{38,63\} & 8 & \frac{512}{567 \sqrt{7}} & \frac{1}{16}
   \left(7+4 \sqrt{7} \csc
   ^{-1}\left(\frac{4}{\sqrt{7}}\right)\right) & 0.915544
   \\
 \{40,63\} & 8 & \frac{512}{567 \sqrt{7}} &
   -\frac{49}{192}+\frac{\sqrt{14}}{3} & 0.992011 \\
 \{48,49\} & 8 & \frac{196 \sqrt{2}}{729} & \frac{3
   \left(22 \sqrt{6}+49 \cos
   ^{-1}\left(\frac{5}{7}\right)\right)}{112 \sqrt{7}} &
   0.930129 \\
 \{49,63\} & 8 & \frac{128 \pi }{567} & \frac{\sqrt{7}+4
   \csc ^{-1}\left(\frac{4}{\sqrt{7}}\right)}{4 \pi } &
   0.440596 \\
 \{53,63\} & 8 & \frac{128 \pi }{567} & \frac{19
   \sqrt{7}+48 \csc
   ^{-1}\left(\frac{4}{\sqrt{7}}\right)}{48 \pi } &
   0.563412 \\
 \{63,64\} & 8 & \frac{512 \sqrt{2}}{1701} & 1-\frac{7
   \sqrt{\frac{7}{2}}}{64} & 0.795378 \\
 \{63,66\} & 8 & \frac{512 \sqrt{2}}{1701} & \frac{19
   \sqrt{7}+48 \cos ^{-1}\left(\frac{3}{4}\right)}{64
   \sqrt{2}} & 0.938690
\end{array}
\right).
\end{displaymath}
\subsection{Boundary States}
Here, we have for the one-dimensional sets of (exterior) boundary states
the results
\begin{equation}
\left(
\begin{array}{lllll}
 \{1,13\} & \frac{4}{9} & 0 & 0 & 0. \\
 \{3,11\} & \frac{2}{3} & \frac{2}{9} & \frac{1}{3} &
   0.333333 \\
 \{3,13\} & \frac{4}{9} & \frac{2}{9} & \frac{1}{2} & 0.5
   \\
 \{3,25\} & \frac{2}{3} & \frac{2}{9} & \frac{1}{3} &
   0.333333 \\
 \{8,13\} & \frac{4}{3 \sqrt{3}} & \frac{1}{3 \sqrt{3}} &
   \frac{1}{4} & 0.25 \\
 \{8,25\} & \frac{1}{\sqrt{3}} & \frac{2}{3 \sqrt{3}} &
   \frac{2}{3} & 0.666667 \\
 \{11,15\} & \frac{8}{9 \sqrt{3}} & \frac{4}{9} &
   \frac{\sqrt{3}}{2} & 0.866025 \\
 \{11,24\} & \frac{5}{18} & \frac{2}{9} & \frac{4}{5} &
   0.8 \\
 \{13,24\} & \frac{5}{9} & \frac{2 \sqrt{5}}{9} &
   \frac{2}{\sqrt{5}} & 0.894427 \\
 \{13,35\} & \frac{4}{15} & \frac{2}{9} & \frac{5}{6} &
   0.833333 \\
 \{15,16\} & \frac{8 \sqrt{\frac{2}{3}}}{9} & \frac{2
   \sqrt{\frac{2}{3}}}{3} & \frac{3}{4} & 0.75 \\
 \{15,44\} & \frac{8 \sqrt{\frac{2}{3}}}{9} & \frac{2
   \sqrt{\frac{2}{3}}}{3} & \frac{3}{4} & 0.75 \\
 \{15,55\} & \frac{10 \sqrt{\frac{2}{3}}}{9} & \frac{2
   \sqrt{\frac{2}{3}}}{9} & \frac{1}{5} & 0.2 \\
 \{16,24\} & \frac{5}{9} & \frac{4}{9} & \frac{4}{5} & 0.8
   \\
 \{20,24\} & \frac{5}{9} & \frac{4}{9} & \frac{4}{5} & 0.8
   \\
 \{24,25\} & \frac{5 \sqrt{\frac{5}{2}}}{9} &
   \frac{\sqrt{\frac{5}{2}}}{9} & \frac{1}{5} & 0.2 \\
 \{24,27\} & \frac{5 \sqrt{\frac{5}{2}}}{9} & \frac{2
   \sqrt{10}}{9} & \frac{4}{5} & 0.8
\end{array}
\right)
\end{equation}
\begin{displaymath}
\left(
\begin{array}{lllll}
 \{24,44\} & \frac{\sqrt{10}}{3} &
   \frac{\sqrt{\frac{5}{2}}}{9} & \frac{1}{6} & 0.166667
   \\
 \{24,46\} & \frac{5 \sqrt{\frac{5}{2}}}{9} &
   \frac{\sqrt{\frac{5}{2}}}{9} & \frac{1}{5} & 0.2 \\
 \{24,55\} & \frac{5 \sqrt{\frac{5}{2}}}{9} & \frac{2
   \sqrt{10}}{9} & \frac{4}{5} & 0.8 \\
 \{24,59\} & \frac{5 \sqrt{\frac{5}{2}}}{9} & \frac{2
   \sqrt{10}}{9} & \frac{4}{5} & 0.8 \\
 \{24,70\} & \frac{5 \sqrt{\frac{5}{2}}}{9} & \frac{2
   \sqrt{10}}{9} & \frac{4}{5} & 0.8 \\
 \{24,72\} & \frac{\sqrt{10}}{3} &
   \frac{\sqrt{\frac{5}{2}}}{9} & \frac{1}{6} & 0.166667
   \\
 \{25,35\} & \frac{4}{3 \sqrt{5}} & \frac{4}{9} &
   \frac{\sqrt{5}}{3} & 0.745356 \\
 \{35,46\} & \frac{14}{3 \sqrt{15}} & \frac{2}{3
   \sqrt{15}} & \frac{1}{7} & 0.142857 \\
 \{35,70\} & \frac{4}{\sqrt{15}} & \frac{2
   \sqrt{\frac{5}{3}}}{3} & \frac{5}{6} & 0.833333 \\
 \{38,48\} & \frac{7 \sqrt{\frac{2}{3}}}{9} & \frac{4}{9}
   & \frac{2 \sqrt{6}}{7} & 0.699854 \\
 \{38,63\} & \frac{16}{63} & \frac{2}{9} & \frac{7}{8} &
   0.875 \\
 \{40,63\} & \frac{32}{63} & \frac{8
   \sqrt{\frac{2}{7}}}{9} & \frac{\sqrt{\frac{7}{2}}}{2} &
   0.935414 \\
 \{48,49\} & \frac{7 \sqrt{\frac{7}{3}}}{9} & \frac{2
   \sqrt{\frac{7}{3}}}{3} & \frac{6}{7} & 0.857143 \\
 \{49,63\} & \frac{16}{9 \sqrt{7}} & \frac{4}{9} &
   \frac{\sqrt{7}}{4} & 0.661438 \\
 \{53,63\} & \frac{16}{9 \sqrt{7}} & \frac{4}{9} &
   \frac{\sqrt{7}}{4} & 0.661438 \\
 \{63,64\} & \frac{32}{9 \sqrt{7}} & \frac{4}{9 \sqrt{7}}
   & \frac{1}{8} & 0.125 \\
 \{63,66\} & \frac{32}{9 \sqrt{7}} & \frac{4 \sqrt{7}}{9}
   & \frac{7}{8} & 0.875
\end{array}
\right).
\end{displaymath}
\subsection{Length of Separability-Nonseparability {\it Interior} Boundary}
In the following arrays, we present the HS-length 
(and now a numerical approximation to it) of the common border
separating the states lacking a PPR from those that possess a PPT
for each specific scenario.
\begin{equation}
\left(
\begin{array}{lll}
 \{1,13\} & \frac{4}{9} & 0.444444 \\
 \{3,11\} & \frac{4}{9 \sqrt{3}} & 0.2566 \\
 \{3,13\} & \frac{4}{9} & 0.444444 \\
 \{3,25\} & \frac{4}{9 \sqrt{3}} & 0.2566 \\
 \{8,13\} & \frac{2}{27} \left(-2 \sqrt{3}+3
   \sqrt{6}\right) & 0.287731 \\
 \{8,25\} & \frac{1}{297} \left(11 \sqrt{3}+12
   \sqrt{33}\right) & 0.296254 \\
 \{11,15\} & \frac{4}{9} & 0.444444 \\
 \{11,24\} & \frac{4}{11} & 0.363636 \\
 \{13,24\} & \frac{4}{11} & 0.363636 \\
 \{13,35\} & \frac{7}{18} & 0.388889 \\
 \{15,16\} & \frac{1}{621} \left(23 \sqrt{6}+12
   \sqrt{138}\right) & 0.317724 \\
 \{15,44\} & \frac{2}{27} \left(3+\sqrt{6}\right) &
   0.403666 \\
 \{15,55\} & \frac{1}{45} \left(-5 \sqrt{6}+8
   \sqrt{10}\right) & 0.290017 \\
 \{16,24\} & \frac{4}{9} & 0.444444 \\
 \{20,24\} & \frac{4}{9} & 0.444444 \\
 \{24,25\} & \frac{117 \sqrt{10}+40 \sqrt{390}}{3510} &
   0.330462 \\
 \{24,27\} & \frac{117 \sqrt{10}+40 \sqrt{390}}{3510} &
   0.330462
\end{array}
\right)
\end{equation}
\begin{displaymath}
\left(
\begin{array}{lll}
 \{24,44\} & -\frac{4}{135} \left(3 \sqrt{10}-5
   \sqrt{15}\right) & 0.292684 \\
 \{24,46\} & \frac{1}{18} \left(4+\sqrt{10}\right) &
   0.397904 \\
 \{24,55\} & \frac{117 \sqrt{10}+40 \sqrt{390}}{3510} &
   0.330462 \\
 \{24,59\} & \frac{1}{18} \left(4+\sqrt{10}\right) &
   0.397904 \\
 \{24,70\} & \frac{117 \sqrt{10}+40 \sqrt{390}}{3510} &
   0.330462 \\
 \{24,72\} & -\frac{4}{135} \left(3 \sqrt{10}-5
   \sqrt{15}\right) & 0.292684 \\
 \{25,35\} & \frac{4}{9} & 0.444444 \\
 \{35,46\} & -\frac{2}{189} \left(7 \sqrt{15}-12
   \sqrt{21}\right) & 0.295027 \\
 \{35,70\} & \frac{4 \left(59 \sqrt{15}+15
   \sqrt{885}\right)}{7965} & 0.338853 \\
 \{38,48\} & \frac{4}{9} & 0.444444 \\
 \{38,63\} & \frac{12}{29} & 0.413793 \\
 \{40,63\} & \frac{12}{29} & 0.413793 \\
 \{48,49\} & \frac{415 \sqrt{21}+84 \sqrt{1743}}{15687} &
   0.344789 \\
 \{49,63\} & \frac{4}{9} & 0.444444 \\
 \{53,63\} & \frac{4}{9} & 0.444444 \\
 \{63,64\} & \frac{333 \sqrt{7}+56 \sqrt{777}}{6993} &
   0.349209 \\
 \{63,66\} & \frac{333 \sqrt{7}+56 \sqrt{777}}{6993} &
   0.349209
\end{array}
\right).
\end{displaymath}
\section{The case $n=10$} \label{secN=10}
\subsection{$5 \times 2$ Decomposition}
We first compute the partial transpose by transposing in place the ($2^2$)
four $5 \times 5$ blocks of our 
set of two-dimensional $10 \times 10$ density matrices. Our analysis yielded
for the HS-total volumes and PPT-probabilities
\begin{equation} \label{first10}
\left(
\begin{array}{lllll}
 \{1,29\} & 480 & \frac{4}{25} & \frac{\pi }{4} & 0.785398
   \\
 \{3,27\} & 4 & \frac{8 \sqrt{2}}{75} & \frac{1}{\sqrt{2}}
   & 0.707107 \\
 \{3,29\} & 12 & \frac{4}{25} & \frac{5}{6} & 0.833333 \\
 \{3,49\} & 12 & \frac{8 \sqrt{2}}{75} & \frac{5}{4
   \sqrt{2}} & 0.883883 \\
 \{8,29\} & 4 & \frac{6}{25} & \frac{1}{\sqrt{3}} &
   0.577350 \\
 \{8,31\} & 8 & \frac{3 \sqrt{3}}{25} & \frac{7}{9} &
   0.777778 \\
 \{8,49\} & 4 & \frac{3 \sqrt{2}}{25} & \sqrt{\frac{2}{3}}
   & 0.816497 \\
 \{8,55\} & 4 & \frac{3 \sqrt{3}}{25} & \frac{8}{9} &
   0.888889 \\
 \{8,64\} & 8 & \frac{3 \sqrt{2}}{25} & \frac{7}{3
   \sqrt{6}} & 0.952579 \\
 \{8,68\} & 4 & \frac{6}{25} & \frac{4}{3 \sqrt{3}} &
   0.769800 \\
 \{15,31\} & 6 & \frac{32 \sqrt{\frac{2}{3}}}{75} &
   \frac{1}{2} & 0.500000 \\
 \{15,33\} & 6 & \frac{8 \sqrt{\frac{2}{3}}}{25} &
   \frac{3}{4} & 0.750000 \\
 \{15,64\} & 6 & \frac{32 \sqrt{2}}{225} &
   \frac{\sqrt{3}}{2} & 0.866025 \\
 \{15,72\} & 2 & \frac{8 \sqrt{\frac{2}{3}}}{25} &
   \frac{11}{12} & 0.916667 \\
 \{15,81\} & 6 & \frac{32 \sqrt{2}}{225} & \frac{9
   \sqrt{3}}{16} & 0.974279 \\
 \{15,87\} & 2 & \frac{32 \sqrt{\frac{2}{3}}}{75} &
   \frac{11}{16} & 0.687500 \\
 \{24,33\} & 8 & \frac{\sqrt{2}}{3} & \frac{1}{\sqrt{5}} &
   0.447214 \\
 \{24,81\} & 8 & \frac{1}{3 \sqrt{2}} & \frac{2}{\sqrt{5}}
   & 0.894427 \\
 \{27,35\} & 8 & \frac{9 \sqrt{3} \pi }{125} & \frac{14
   \sqrt{5}+27 \csc
   ^{-1}\left(\frac{3}{\sqrt{5}}\right)}{27 \pi } &
   0.636783 \\
 \{27,48\} & 2 & \frac{49 \sqrt{\frac{7}{3}}}{300} &
   \frac{1}{28} \left(12+7 \sqrt{6} \sin
   ^{-1}\left(\frac{2 \sqrt{6}}{7}\right)\right) &
   0.903278
\end{array}
\right)
\end{equation}
\begin{displaymath}
\left(
\begin{array}{lllll}
 \{29,48\} & 6 & \frac{49 \sqrt{\frac{7}{3}}}{300} &
   \frac{4}{147} \left(-9+7 \sqrt{42}\right) & 0.989529 \\
 \{29,63\} & 4 & \frac{128}{175 \sqrt{7}} & \frac{1}{16}
   \left(7+4 \sqrt{7} \csc
   ^{-1}\left(\frac{4}{\sqrt{7}}\right)\right) & 0.915544
   \\
 \{31,63\} & 8 & \frac{128}{175 \sqrt{7}} &
   -\frac{49}{192}+\frac{\sqrt{14}}{3} & 0.992011 \\
 \{31,80\} & 6 & \frac{243}{800} & \frac{4}{9}+\frac{\sin
   ^{-1}\left(\frac{4 \sqrt{2}}{9}\right)}{\sqrt{2}} &
   0.925046 \\
 \{33,80\} & 6 & \frac{243}{800} & \frac{8}{243}
   \left(-8+27 \sqrt{2}\right) & 0.993704 \\
 \{33,99\} & 8 & \frac{4 \sqrt{5}}{27} & \frac{3}{20}
   \left(3+5 \sin ^{-1}\left(\frac{3}{5}\right)\right) &
   0.932626 \\
 \{35,36\} & 8 & \frac{24 \sqrt{2}}{125} & \frac{14
   \sqrt{5}+27 \cos ^{-1}\left(\frac{2}{3}\right)}{24
   \sqrt{6}} & 0.918793 \\
 \{36,48\} & 2 & \frac{49}{600} \sqrt{\frac{7}{2}} \pi  &
   \frac{2 \sqrt{6}+7 \sin ^{-1}\left(\frac{2
   \sqrt{6}}{7}\right)}{7 \pi } & 0.469522 \\
\end{array} .
\right) 
\end{displaymath}
\subsection{$2 \times 5$ Decomposition}
Now, we compute the partial transpose by transposing in place the ($5^2$)
twenty-five $4 \times 4$ blocks of the $10 \times 10$ density matrices.
\begin{equation}
\left(
\begin{array}{lllll}
 \{3,6\} & 16 & \frac{8 \sqrt{2}}{75} & \frac{1}{\sqrt{2}}
   & 0.707107 \\
 \{4,18\} & 480 & \frac{4}{25} & \frac{\pi }{4} & 0.785398
   \\
 \{6,8\} & 2 & \frac{9}{200} \sqrt{\frac{3}{2}} \pi  &
   \frac{26 \sqrt{2}+27 \tan ^{-1}\left(2
   \sqrt{2}\right)}{27 \pi } & 0.825312 \\
 \{6,15\} & 2 & \frac{16 \sqrt{\frac{2}{3}}}{75} &
   \frac{1}{24} \left(9+2 \sqrt{3} \pi \right) & 0.828450
   \\
 \{8,9\} & 2 & \frac{3 \sqrt{2}}{25} & \frac{52+27
   \sqrt{2} \sec ^{-1}(3)}{48 \sqrt{6}} & 0.842035 \\
 \{8,22\} & 6 & \frac{3 \sqrt{3}}{25} & \frac{8}{9} &
   0.888889 \\
 \{8,29\} & 6 & \frac{6}{25} & \frac{4}{3 \sqrt{3}} &
   0.769800 \\
 \{9,15\} & 2 & \frac{4 \sqrt{2} \pi }{75} &
   \frac{1}{3}+\frac{\sqrt{3}}{2 \pi } & 0.608998 \\
 \{15,22\} & 6 & \frac{32 \sqrt{\frac{2}{3}}}{75} &
   \frac{1}{2} & 0.500000 \\
 \{15,29\} & 6 & \frac{32 \sqrt{2}}{225} &
   \frac{\sqrt{3}}{2} & 0.866025 \\
 \{18,24\} & 4 & \frac{1}{16} \sqrt{\frac{5}{2}} \pi  &
   \frac{92+75 \sin ^{-1}\left(\frac{4}{5}\right)}{75 \pi
   } & 0.685627 \\
 \{18,35\} & 4 & \frac{36 \sqrt{\frac{3}{5}}}{125} &
   \frac{1}{12} \left(5+3 \sqrt{5} \csc
   ^{-1}\left(\frac{3}{\sqrt{5}}\right)\right) & 0.886838
   \\
 \{24,25\} & 4 & \frac{1}{3 \sqrt{2}} & \frac{92+75 \cos
   ^{-1}\left(\frac{3}{5}\right)}{80 \sqrt{5}} & 0.903076
   \\
 \{24,46\} & 4 & \frac{1}{\sqrt{10}} & \frac{14}{15} &
   0.933333 \\
 \{24,57\} & 4 & \frac{\sqrt{2}}{3} & \frac{7}{5 \sqrt{5}}
   & 0.626099
\end{array}
\right)
\end{equation}
\begin{displaymath}
\left(
\begin{array}{lllll}
 \{25,35\} & 4 & \frac{9 \sqrt{3} \pi }{125} &
   \frac{\sqrt{5}+3 \csc
   ^{-1}\left(\frac{3}{\sqrt{5}}\right)}{3 \pi } &
   0.504975 \\
 \{35,46\} & 4 & \frac{24 \sqrt{\frac{2}{5}}}{25} &
   \frac{1}{\sqrt{6}} & 0.408248 \\
 \{35,57\} & 4 & \frac{24 \sqrt{2}}{125} &
   \sqrt{\frac{5}{6}} & 0.912871 \\
 \{38,48\} & 6 & \frac{49}{600} \sqrt{\frac{7}{2}} \pi  &
   \frac{22 \sqrt{6}+49 \sin ^{-1}\left(\frac{2
   \sqrt{6}}{7}\right)}{49 \pi } & 0.596820 \\
 \{38,63\} & 6 & \frac{128}{175 \sqrt{7}} & \frac{1}{16}
   \left(7+4 \sqrt{7} \csc
   ^{-1}\left(\frac{4}{\sqrt{7}}\right)\right) & 0.915544
   \\
 \{48,49\} & 6 & \frac{49 \sqrt{2}}{225} & \frac{3
   \left(22 \sqrt{6}+49 \cos
   ^{-1}\left(\frac{5}{7}\right)\right)}{112 \sqrt{7}} &
   0.930129 \\
 \{48,78\} & 2 & \frac{7 \sqrt{\frac{7}{3}}}{25} &
   \frac{20}{21} & 0.952381 \\
 \{48,93\} & 2 & \frac{98}{75 \sqrt{3}} & \frac{10}{7
   \sqrt{7}} & 0.539949 \\
 \{49,63\} & 6 & \frac{32 \pi }{175} & \frac{\sqrt{7}+4
   \csc ^{-1}\left(\frac{4}{\sqrt{7}}\right)}{4 \pi } &
   0.440596 \\
 \{63,78\} & 2 & \frac{128 \sqrt{\frac{2}{7}}}{75} &
   \frac{1}{2 \sqrt{2}} & 0.353553 \\
 \{63,93\} & 2 & \frac{128 \sqrt{2}}{525} &
   \frac{\sqrt{\frac{7}{2}}}{2} & 0.935414 \\
 \{66,80\} & 8 & \frac{243 \pi }{800 \sqrt{2}} &
   \frac{\frac{172 \sqrt{2}}{243}+\sin ^{-1}\left(\frac{4
   \sqrt{2}}{9}\right)}{\pi } & 0.534977 \\
 \{66,99\} & 8 & \frac{4 \sqrt{5}}{27} & \frac{3}{20}
   \left(3+5 \sin ^{-1}\left(\frac{3}{5}\right)\right) &
   0.932626 \\
 \{80,81\} & 8 & \frac{27}{50 \sqrt{2}} & \frac{43}{54
   \sqrt{2}}+\frac{9}{16} \cos
   ^{-1}\left(\frac{7}{9}\right) & 0.945383 \\
 \{81,99\} & 8 & \frac{\sqrt{5} \pi }{9} & \frac{3+5 \sin
   ^{-1}\left(\frac{3}{5}\right)}{5 \pi } & 0.395819
\end{array}
\right).
\end{displaymath}
\subsection{Bi-PPT}
Now, we require that the $10 \times 10$ density matrices be positive
under {\it both} the forms of partial transposition employed immediately above.
We obtained the (extensive) results
\begin{equation}
\left(
\begin{array}{lllll}
 \{1,29\} & 904 & \frac{4}{25} & \frac{\pi }{4} & 0.785398
   \\
 \{3,6\} & 20 & \frac{8 \sqrt{2}}{75} & \frac{1}{\sqrt{2}}
   & 0.707107 \\
 \{3,29\} & 12 & \frac{4}{25} & \frac{5}{6} & 0.833333 \\
 \{3,51\} & 6 & \frac{8 \sqrt{2}}{75} & \frac{5}{4
   \sqrt{2}} & 0.883883 \\
 \{6,8\} & 2 & \frac{9}{200} \sqrt{\frac{3}{2}} \pi  &
   \frac{26 \sqrt{2}+27 \tan ^{-1}\left(2
   \sqrt{2}\right)}{27 \pi } & 0.825312 \\
 \{6,15\} & 2 & \frac{16 \sqrt{\frac{2}{3}}}{75} &
   \frac{1}{24} \left(9+2 \sqrt{3} \pi \right) & 0.828450
   \\
 \{8,9\} & 2 & \frac{3 \sqrt{2}}{25} & \frac{52+27
   \sqrt{2} \sec ^{-1}(3)}{48 \sqrt{6}} & 0.842035 \\
 \{8,22\} & 8 & \frac{3 \sqrt{3}}{25} & \frac{8}{9} &
   0.888889 \\
 \{8,29\} & 4 & \frac{6}{25} & \frac{1}{\sqrt{3}} &
   0.577350 \\
 \{8,31\} & 6 & \frac{3 \sqrt{3}}{25} & \frac{7}{9} &
   0.777778 \\
 \{8,42\} & 2 & \frac{3 \sqrt{3}}{25} & \frac{2}{3} &
   0.666667 \\
 \{8,49\} & 4 & \frac{3 \sqrt{2}}{25} & \sqrt{\frac{2}{3}}
   & 0.816497 \\
 \{8,53\} & 6 & \frac{6}{25} & \frac{4}{3 \sqrt{3}} &
   0.769800 \\
 \{8,64\} & 8 & \frac{3 \sqrt{2}}{25} & \frac{7}{3
   \sqrt{6}} & 0.952579 \\
 \{9,15\} & 2 & \frac{4 \sqrt{2} \pi }{75} &
   \frac{1}{3}+\frac{\sqrt{3}}{2 \pi } & 0.608998 \\
 \{15,22\} & 10 & \frac{32 \sqrt{\frac{2}{3}}}{75} &
   \frac{1}{2} & 0.500000 \\
 \{15,29\} & 12 & \frac{32 \sqrt{2}}{225} &
   \frac{\sqrt{3}}{2} & 0.866025 \\
 \{15,33\} & 6 & \frac{8 \sqrt{\frac{2}{3}}}{25} &
   \frac{3}{4} & 0.750000 \\
 \{15,72\} & 2 & \frac{8 \sqrt{\frac{2}{3}}}{25} &
   \frac{11}{12} & 0.916667 \\
 \{15,81\} & 4 & \frac{32 \sqrt{2}}{225} & \frac{9
   \sqrt{3}}{16} & 0.974279
\end{array}
\right)
\end{equation}
\begin{displaymath}
\left(
\begin{array}{lllll}
 \{15,87\} & 2 & \frac{32 \sqrt{\frac{2}{3}}}{75} &
   \frac{11}{16} & 0.687500 \\
 \{18,24\} & 4 & \frac{1}{16} \sqrt{\frac{5}{2}} \pi  &
   \frac{92+75 \sin ^{-1}\left(\frac{4}{5}\right)}{75 \pi
   } & 0.685627 \\
 \{18,35\} & 4 & \frac{36 \sqrt{\frac{3}{5}}}{125} &
   \frac{1}{12} \left(5+3 \sqrt{5} \csc
   ^{-1}\left(\frac{3}{\sqrt{5}}\right)\right) & 0.886838
   \\
 \{24,25\} & 4 & \frac{1}{3 \sqrt{2}} & \frac{92+75 \cos
   ^{-1}\left(\frac{3}{5}\right)}{80 \sqrt{5}} & 0.903076
   \\
 \{24,33\} & 8 & \frac{\sqrt{2}}{3} & \frac{1}{\sqrt{5}} &
   0.447214 \\
 \{24,46\} & 4 & \frac{1}{\sqrt{10}} & \frac{14}{15} &
   0.933333 \\
 \{24,81\} & 8 & \frac{1}{3 \sqrt{2}} & \frac{2}{\sqrt{5}}
   & 0.894427 \\
 \{24,89\} & 2 & \frac{\sqrt{2}}{3} & \frac{7}{5 \sqrt{5}}
   & 0.626099 \\
\end{array}
\right)
\end{displaymath}
\begin{displaymath}
\left(
\begin{array}{lllll}
 \{36,48\} & 2 & \frac{49}{600} \sqrt{\frac{7}{2}} \pi  &
   \frac{2 \sqrt{6}+7 \sin ^{-1}\left(\frac{2
   \sqrt{6}}{7}\right)}{7 \pi } & 0.469522 \\
 \{38,48\} & 10 & \frac{49}{600} \sqrt{\frac{7}{2}} \pi  &
   \frac{22 \sqrt{6}+49 \sin ^{-1}\left(\frac{2
   \sqrt{6}}{7}\right)}{49 \pi } & 0.596820 \\
 \{48,49\} & 2 & \frac{49 \sqrt{2}}{225} & \frac{3 \left(2
   \sqrt{6}+7 \cos
   ^{-1}\left(\frac{5}{7}\right)\right)}{16 \sqrt{7}} &
   0.731739 \\
 \{48,51\} & 10 & \frac{49 \sqrt{2}}{225} & \frac{3
   \left(22 \sqrt{6}+49 \cos
   ^{-1}\left(\frac{5}{7}\right)\right)}{112 \sqrt{7}} &
   0.930129 \\
 \{48,64\} & 4 & \frac{49 \sqrt{2}}{225} & 1-\frac{3
   \sqrt{\frac{3}{14}}}{7} & 0.801610 \\
 \{48,78\} & 2 & \frac{7 \sqrt{\frac{7}{3}}}{25} &
   \frac{20}{21} & 0.952381 \\
 \{48,93\} & 2 & \frac{98}{75 \sqrt{3}} & \frac{10}{7
   \sqrt{7}} & 0.539949 \\
 \{49,63\} & 8 & \frac{32 \pi }{175} & \frac{\sqrt{7}+4
   \csc ^{-1}\left(\frac{4}{\sqrt{7}}\right)}{4 \pi } &
   0.440596 \\
\end{array} .
\right)
\end{displaymath}
The probability 0.993704 (corrresponding to the pair of Lie generators 
numbered $\{33,80 \}$) is the largest of any recorded in 
all our results above. (This also occurs in (\ref{first10}).)
\section{Analyses of Scenarios with {\it More} than Two Parameters} \label{finalSEC}
We found it considerably 
simpler to extend the Jak\'obczyk-Siennicki model \cite{jak} from
two-qubit systems ($n=4$) to higher $n$ --- as illustrated above --- than 
to extend it from
two-dimensional sections of Bloch vectors to $m$-dimensional sections
($m \ge 3$), even just for the case $n=4$. (However, 
we were able to obtain a highly interesting set
of {\it exact} HS separability probabilites for certain $m=3,n=4$ systems, 
using the Jaynes maximum-entropy 
principle in conjunction with the new integration over implicitly defined
regions 
feature of Mathematica, in \cite[Fig.~11]{slaterjpanew}.) 
{\it A fortiori}, it appears that
the determination of the HS separable volume of  the {\it fifteen}-dimensional
convex set of $4 \times 4$ density matrices --- conjectured 
on the basis of an extensive quasi-Monte Carlo analysis in to be
$(3^3 5^7 \sqrt{3})^{-1} \approx 2.73707 \cdot 10^{-7} $  \cite[eq. (41)]{slaterPRA} --- would  have to proceed in some
quite different analytic fashion to that pursued here.
(Based on our experience in the above-reported analyses, it appears to be
a necessary condition for obtaining exact HS separability/PPT-probabilities
that explicit formulas be available for the eigenvectors of both the
class of density matrices under consideration {\it and} of their partial
transposes.)
\subsection{$m=3,n=4$}
We have been able to find, up to this point in time, that 
for the {\it three}-dimensional two-qubit ($m=3, 
n=4$) scenarios generated by
the four {\it triads} of Gell-Mann matrices $\{1,4,6\}$, $\{1,5,7\}$  
$\{2,4,7\}$ and $\{2,5,6\}$, 
the volume of separable
states is --- having to resort to numerical methods --- 0.478512 
and of all the (separable and nonseparable/entangled)
states, 0.61685, for a separability probability of 0.775734.
For the scenario $\{10,12,13\}$, the separable volume remains the same, 
but the total volume is {\it exactly} $\frac{\pi}{6} \approx 0.523599$
for an HS separability probability of 0.913891.
\subsection{Two-Dimensional Boundaries of $m=3,n=4$ Systems}
Of course, it we restrict attention to the generic 
boundary states of the three-dimensional scenarios, we only have to perform two-dimensional computations.
Thus, we were able to find that for the {\it triadic} 
scenarios $\{1,3,6\},\{1,3,7\},
\{1,3,9\}, \{1,3,10\}, \{1,4,9\}$ and $\{1,5,10\}$, amongst others,  the HS-area 
of the states with degenerate spectra is $\frac{\pi}{8}$
and that of the separable component of this area, one-half that value.
For the triadic scenarios $\{1,4,6\}$ and $\{1,5,7\}$, 
the separable component of the boundary states
has area $\frac{\pi}{4}$ and the total area is $\frac{1}{2} 
\Big(\sqrt{5} + 6 \sin^{-1}(\frac{1}{\sqrt{6}})\Big)$ for a separability probability of 0.165025.
Also, for several scenarios (for instance, $\{3,4,9\}$), we have a total area
of $\frac{2 \sqrt{2}}{3}$, a separable area of $\frac{1}{3}$, giving a 
separability probability of $\frac{1}{2 \sqrt{2}} \approx 0.353553$.

Now, we present all our results of this type (two-dimensional exterior 
boundaries of three-dimensional scenarios) in the following array.
The first column gives the corresponding {\it triad} of Gell-Mann matrices, the second column shows the total HS-area of the boundary states, the third column gives the exact separability probability, and the last, a numerical approximation
to the probability. (There may exist additional nontrivial scenarios, as we 
were not readily able to fully analyze all $2730 = 13 \cdot 14 \cdot 15$ 
possible triads of $4 \times 4$ Gell-Mann matrices.)
\begin{equation}
\left(
\begin{array}{llll}
 \{1,3,6\} & \frac{\pi }{4} & \frac{1}{2} & 0.500000 \\
 \{1,4,6\} & \frac{1}{2} \left(\sqrt{5}+6 \sin
   ^{-1}\left(\frac{1}{\sqrt{6}}\right)\right) & \frac{\pi
   }{4 \sqrt{5}+24 \csc ^{-1}\left(\sqrt{6}\right)} &
   0.165025 \\
 \{3,4,6\} & \frac{1}{2} \left(\sqrt{5}+6 \sin
   ^{-1}\left(\frac{1}{\sqrt{6}}\right)\right) &
   \frac{1}{2}+\frac{4 \left(-1+\sqrt{2}\right)}{3
   \left(\sqrt{5}+6 \csc
   ^{-1}\left(\sqrt{6}\right)\right)} & 0.616044 \\
 \{3,4,9\} & \frac{2 \sqrt{2}}{3} & \frac{1}{2 \sqrt{2}} &
   0.353553 \\
 \{3,6,7\} & \frac{1}{2} \left(\sqrt{5}+6 \sin
   ^{-1}\left(\frac{1}{\sqrt{6}}\right)\right) &
   \frac{1}{2} & 0.500000 \\
 \{3,6,8\} & \frac{3 \pi }{2} & -\frac{\sqrt{15}-8 \pi +8
   \tan ^{-1}\left(\sqrt{\frac{3}{5}}\right)}{8 \pi } &
   0.636114 \\
 \{6,9,15\} & \frac{3 \pi }{4} & \frac{4-2 \sqrt{5}+3 \pi
   -12 \csc ^{-1}\left(\sqrt{6}\right)}{6 \pi } & 0.207232
   \\
 \{8,9,10\} & \frac{3}{16} \left(4+\sqrt{7}+2 \pi +8 \cot
   ^{-1}\left(\sqrt{7}\right)\right) & \frac{2 (2+\pi
   )}{4+\sqrt{7}+2 \pi +8 \cot ^{-1}\left(\sqrt{7}\right)}
   & 0.650017 \\
 \{9,11,13\} & \frac{3 \pi }{2} & \frac{1}{12} & 0.0833333
\end{array}
\right).
\end{equation}
\subsection{9- and 15-Parameter Analyses ($m=9,15,n=4$)} \label{final}
Now, we sought to make some progress in obtaining the 
(conjecturally exact) Hilbert-Schmidt
volume of the separable $4 \times 4$ density matrices, in both the 
9-dimensional case of real density matrices and the 15-dimensional case
of (fully general) complex density matrices.
In both cases, we dispensed with the Bloch vector parameterization 
\cite{kk,kimura,byrd} 
used in the above analyses (neither did we employ the integration 
over implicitly defined regions 
capabilities of Mathematica version 5.1), and adopted a simple, naive
parameterization, in which the four diagonal elements 
of the density matrices were denoted
$a,b,c,1-a-b-c$ and the off-diagonal (upper triangular) 
elements, $\alpha_{ij}+ i 
\beta_{ij}$ (in the real $m =9$ 
case, of course, all $\beta$'s equal zero).
(In order to compare our results here with the HS-volume formulas
of \.Zyczkowski and Sommers \cite{hilb2}, the volumes we do report below
are our initial volumes multiplied 
by factors of $2^7$ in the complex case, and $2^4$ 
in the real case. In all our earlier analyses above, we have simply taken
the HS-volume element to equal 1.)

In both ($m=9,15$) of these cases, we pursued the same analytical strategy.
We required that the six principal $2 \times 2$ minors of the density
matrices and/or their partial transposes have nonnegative determinants.
This is (only) one of the requirements for 
nonnegative-definiteness (cf. 
\cite[eq. (12)]{bloore}).
Ideally, we would also have required that the leading principal $3 \times 
3$ minor have nonnegative determinant and also that the determinant of
the matrix be nonnegative. But these last two requirements were too 
computationally onerous to impose (at least in our first round of efforts). So, our analytical strategy should yield {\it upper}
bounds on the Hilbert-Schmidt volumes in these cases.
\subsubsection{9-Dimensional Real Case}
When we only required that the six principal $2 \times 2$ 
minors have nonnegative
determinants, we obtained for the volume the result $\frac{\pi^2}{1120} 
\approx 0.00881215$. (We can reduce [improve] this to 
$\frac{\pi^2 (16+\pi^2)}{35840} \approx 0.00712396$ by modifying 
[narrowing], to begin 
with, the integration limits over a {\it single} off-diagonal variable,
so that in addition, to its corresponding $2 \times 2$ minor, a corresponding
$3 \times 3$ minor also has a  nonnegative determinant. 
If we narrow similarly a second set of integration limits, 
this is further reduced
to $\frac{\pi^4}{26880} \approx 0.00362385$. An attempt to add a third set
of similar integration limits --- corresponding to a 
$3 \times 3$ minor --- did not succeed computationally.)
Applying formula (7.7) of the \.Zyczkowski-Sommers study 
\cite{hilb2}, we obtain for the HS-volume of the $4 \times 4$ 
real density matrices, $\frac{\pi^4}{60480} 
\approx 0.0016106$. This is $\frac{\pi^2}{54} \approx 0.18277$ 
times {\it smaller} than our first, principal calculation 
($\frac{\pi^2}{1120}$), so we have a considerable 
overestimation.

If we {\it additionally} imposed the condition that
the six principal minors of the partial transpose also have nonnegative
determinants (only two of them being actually different from the 
original six), the result was $\frac{544}{99225} \approx 0.00548249$.
(Note that $1120 = 2^5 \cdot 5 \cdot 7$ and $99225 = 3^4 \cdot 5^2 \cdot 7^2$.) So (taking the ratio) of this to $\frac{\pi^2}{1120}$, we obtain 
a crude estimate of the HS-separability
probability of the real density matrices is 0.622151.

Unfortunately, our upper bound (0.00548249) 
on the HS-volume of the separable real
two-qubit states is {\it larger} than the (known) HS-volume 
(0.0016106) of the
(separable {\it and} nonseparable) two-qubit states, so we have not yet
succeeded in deriving a nontrivial upper bound on the separable volume.
(The same will be the case in the immediate next analysis.)
\subsubsection{15-Dimensional Complex Case}
Now, when we again required that the six principal $2 \times2$ minors have
nonnegative determinants, we obtained for the corresponding Hilbert-Schmidt
volume $\frac{\pi^6}{7882875} \approx 0.000121959$.
(Note that $7882875 = 3^2 \cdot 5^3 \cdot 7^2 \cdot 11 
\cdot 13$.)
Formula (4.5) of \cite{hilb2} gives us for the 
HS-volume of the 15-dimensional convex set of two-qubit density 
matrices, the value 
$\frac{\pi^6}{85130500} \approx 1.12925 \cdot 10^{-6}$. 
(The ratio of these two volumes --- the measure of our 
overestimation --- is $\frac{7484}{693} \approx 10.7994$.) 
Imposing (just as we did in the {\it real} 9-dimensional case) 
the {\it further} requirements that the six $2 \times 2$ minors
of the partial transpose all have nonnegative determinants, we obtain
a HS-volume of $\frac{1964 \pi^6}{30435780375} \approx 0.0000620378$.
(Observe that 
$30435780375 = 3^5 \cdot 5^3 \cdot 7^2 \cdot 11^2 
\cdot 13^2$.)
So, our crude separability {\it probability} estimate (less than in the 
9-dimensional real case --- as conforms with our intuition) 
is $\frac{1964}{3861} \approx 0.508677$.
Based on certain numerical and theoretical considerations, the actual value
of this (15-dimensional) separability probability has been conjectured 
to be \cite[eq. (43)]{slaterPRA}
\begin{equation}
\frac{2^2 \cdot 3 \cdot 7^2 \cdot 11 \cdot 13 \sqrt{3}}{5^3 \pi^6} 
\approx 0.242379.
\end{equation}
\section{Concluding Remarks}
We have found the newly-introduced capability 
of Mathematica (version 5.1) for integration over implicitly defined 
regions particularly useful for obtaining a very
wide variety of separability (and positive-PPT) probabilities, particularly
for low-dimensional ($m=2,3$) cases, essentially independently of the sizes
($n$) of the corresponding $n \times n$ density matrices analyzed.
The use of such methods for cases $m>=4$ appears, however --- such as 
the two-qubit ($n=4$) scenarios for the real ($m=9$) and 
complex ($m=15$) cases  --- to be particularly 
challenging.

Eggeling and Werner \cite{tilo} studied the separability properties in a 
five-dimensional set of states of quantum systems composed of
{\it three} subsystems of equal but arbitrary finite Hilbert space
dimension. They are the states that commute with unitaries of
the form $U \otimes U \otimes U$.
In \cite{slatertilo}, we evaluated the probabilities of an 
Eggeling-Werner state being biseparable, triseparable or having a 
positive partial transpose with respect to certain partitions. 
However, the Hilbert-Schmidt measure was not employed, but rather
the {\it Bures} one \cite{hanskarol}.

\begin{acknowledgments}
I wish to express gratitude to the Kavli Institute for Theoretical
Physics (KITP)
for computational support in this research 
and to Michael Trott
of Wolfram Research Inc. for
his generous willingness/expertise in assisting with Mathematica computations..
\end{acknowledgments}

\bibliography{Jak3}

\end{document}